\documentclass[structabstract]{aa}  

\usepackage{amsmath, amssymb}
\usepackage{graphicx}
\usepackage{txfonts}
\usepackage{textcomp}
\usepackage{natbib}
\usepackage{wasysym}
\usepackage{hyperref}

\begin{document}
\title{Constraints on rotational mixing from surface evolution
of light elements in massive stars}
\titlerunning{Constraints on rotational mixing}

\author{U. Frischknecht\inst{1}
\and  R. Hirschi\inst{2,3}
\and  G. Meynet\inst{4}
\and  S. Ekstr\"om\inst{4}
\and  C. Georgy\inst{4}
\and  T. Rauscher\inst{1}
\and  C. Winteler\inst{1}
\and  F.-K. Thielemann\inst{1}
}
\authorrunning{U. Frischknecht et al.}


\institute{Department of Physics, University of Basel,
           Klingelbergstrasse 82, 4056 Basel, Switzerland\\
           \email{urs.frischknecht@unibas.ch}
\and
           Astrophysics Group, Keele University, ST5 5BG Keele, UK\\
           \email{hirschi@astro.keele.ac.uk}
\and
           Institute for the Physics and Mathematics of the Universe, 
           University of Tokyo, Kashiwa, 277-8568, Japan 
\and       
           Observatoire Astronomique de l'Universit\'{e} de Gen\`{e}ve, 1290 Sauverny, Switzerland
             }


 
  \abstract
   {Light elements and nitrogen surface abundances together can constrain the mixing efficiencies in massive stars on the main sequence, because moderate mixing in the surface layers leads to depletion of light elements but only later to enrichment in nitrogen.}
   {We want to test the rotational mixing prescriptions included in the Geneva stellar evolution code (GENEC) by following the evolution of surface abundances of light isotopes in massive stars.}
   {The GENEC is a 1D code containing sophisticated prescriptions for rotational mixing. We implemented an extended reaction network into this code including the light elements Li, Be, and B, which allowed us to perform calculations testing the rotation-induced mixing.}
   {We followed 9, 12, and 15~M$_\odot$ models with rotation from the zero age main sequence up to the end of He burning. The calculations show the expected behaviour with faster depletion of light isotopes for faster rotating stars and more massive stars.}
   {We find that the mixing prescriptions used in the present rotating models for massive single stars can account for most of the observations; however, the uncertainties are quite large, making it hard to draw a firm conclusion on the mixing scenario.}


   \maketitle
%

\section{Introduction}

Rotation is beside the stellar mass and the initial chemical composition a key parameter in the evolution of single stars. It affects the physical and chemical structures of the stars and therefore quantities such as lifetime, luminosity, effective temperature $T_{\rm{eff}}$ etc. Recent models including rotation reproduce a wide range of observations better than those without as for example, the nitrogen surface enrichment \citep{2000ApJ...544.1016H,2000A&A...361..101M}, the Wolf-Rayet to O-type star number ratio \citep{2003A&A...404..975M,2005A&A...429..581M,2007ApJ...663..995V}, the ratio of blue to red supergiants in the Small Magellanic Cloud \citep{2001A&A...373..555M}, or the variation with the metallicity of the number ratio of type Ibc to type II supernovae \citep{2009A&A...502..611G}. Still, the treatment of transport of angular momentum and chemical species is thought to be one of the main uncertainties in stellar evolution models. New observational data of late O- and B-type stars from the VLT-FLAMES survey \citep{2005A&A...437..467E, 2006A&A...456..623E, 2007A&A...466..277H, 2007A&A...471..625T} lead to an intense discussion about how well the models with rotation can explain the observed nitrogen surface abundances \citep{2008ApJ...676L..29H,2009A&A...496..841H,2008arXiv0810.0657M} and about whether or not binaries are needed to explain some groups of the observed O- and B-type stars \citep{2008IAUS..250..167L}. Light elements and in particular boron can constrain the mixing induced by rotation and help distinguish between single stars and interacting binaries \citep{2009CoAst.158...55B}. Boron is destroyed at relatively low temperatures ($\approx 6\cdot10^6$~K) where the CNO-cycles are not yet efficient. Therefore modest mixing due to rotation leads to a depletion of light elements at the surface without considerable nitrogen enrichment. This effect can not be explained by mass transfer in a binary system, since there the accreted material is depleted in boron and enriched in nitrogen \citep{1996A&A...308L..13F}. 

Boron is produced in spallation process of CNO atoms in the interstellar medium (ISM) by galactic cosmic rays (GCRs). In massive stars boron is only destroyed. It is the only light element out of Li, Be and B which is observed at the surface of massive  stars (OB-type). Despite the difficulties to measure boron surface abundances an increasing number of boron surface abundances from O- and early B-type stars became available in the last few years \citep{2006ApJ...640.1039M,2002ApJ...565..571V, 2001ApJ...548..429P,1999ApJ...516..342P}. The comparison in \citet{2006ApJ...640.1039M} of observational data with the models of \citet{2000ApJ...544.1016H} shows a good agreement with the exception of three stars (\object{HD 30836}, \object{HD 36591}, \object{HD205021}). The strong boron depletion in these young stars raises the question if the efficiency of surface mixing due to rotation should be stronger or if there are other mixing processes at work. We reexamined this question because the Geneva stellar evolution code (hereafter GENEC) includes the effect of rotation in a different way with respect to the codes which were used in previous works to examine that question. The most important difference comes from the fact that in GENEC the transport of the angular momentum is properly accounted for as an advection process and not as a diffusion process. 

In this paper we present correlations of the surface boron abundances with nitrogen as in \citet{2006ApJ...640.1039M},  \citet{2002ApJ...565..571V} and also with other interesting quantities such as the $^{12}$C$/^{13}$C number ratio or observable physical quantities such as the gravity and the surface velocity. In Sect.~\ref{sec:models} we give a short description of the model ingredients and present the set of simulations performed. In Sect.~\ref{sec:results}, we discuss the results from our models. In Sect.~\ref{sec:obs} we compare them to observations and in Sect.~\ref{sec:conclusion} we summarise the results.  

\section{Stellar model description}
\label{sec:models}
\subsection{Rotation induced mixing}
The Geneva code (GENEC) used for the calculation of our models is described in detail in many previous publications as for example in \citet{2004A&A...425..649H} and more recently \citet{2007Ap&SS.tmp..263E}. Since the mixing efficiency is tested by the light element surface depletion, we will briefly describe here the mixing prescription implemented in GENEC. The horizontal transport of matter is assumed to be much faster than the vertical one, which leads to almost constant angular velocity on isobars. This in turn enables to describe the stellar structure by shellular rotation, which allows to keep the stellar structure equations in one dimension \citep{1992A&A...265..115Z, 1997A&A...321..465M}. The transport of angular momentum in the radiative zones is then described by
\begin{equation}
 \rho\frac{d}{dt}\left(r^2\Omega\right)_{M_r}=\frac{1}{5r^2}\frac{\partial}{\partial r}\left(\rho r^4 \Omega U(r)\right)+\frac{8}{5r^2}\frac{\partial}{\partial r}\left(\rho D_{\rm shear} r^4 \frac{\partial \Omega}{\partial r}\right)
\end{equation}
where $\rho$ is the density, $\Omega$ the angular velocity of a shell, $D_{\rm shear}$ the diffusion coefficient due to the vertical shear turbulence (see Eq.~\ref{eq:dshear} below) and $U(r)$ the quantity intervening in the expression of the radial component of the
meridional velocity which is expressed by $u(r, \theta)=U(r)\cdot P_2(\cos \theta)$. Meridional mixing is an advective process. In contrast with diffusive processes which always smooth gradients, advection can both build up gradients or smooth them. It is therefore  important to account for this process not as a diffusive process but as an advective one. This has been properly done in the present work during the main-sequence (MS) phase. After the MS, the impacts of meridional currents are much less important because the evolutionary timescales become shorter. During the post MS, the main effect governing the evolution
of the angular velocity in the radiative zone is simply the local conservation of the angular momentum.

The transport of chemical composition in the convective core is treated as instantaneous. The size of the convective core is determined by the Schwartzschild criterion to which an overshooting distance $d_{\rm ov}$, which is set by $d_{\rm ov}=\alpha_{\rm ov} \min(H_p,r_{\rm core})$, is added. We adopted $\alpha_{\rm ov}=0.1$ for all calculations. 

The change of the mass fraction, $\dot{X}_i$, of a nuclide $i$ due to rotation induced mixing is described by a diffusion equation (note that nuclear burning and mixing are treated separately).
\begin{equation}
 \left(\frac{dX_i}{dt}\right)_{M_r}= \left(\frac{\partial}{\partial M_r}\right)_t \left[\left(4\pi r^2\rho\right)^2 D_{\rm mix}\left(\frac{\partial X_i}{\partial M_r}\right)_t\right]
\end{equation}
where $D_{\rm{mix}}=D_{\rm{eff}}+D_{\rm{shear}}$ with $D_{\rm{shear}}$ accounting for the vertical mixing introduced by shear turbulence and $D_{\rm eff}$ the diffusion coefficient resulting from the interaction of the strong horizontal mixing induced by shear turbulence and the meridional currents \citep[see][]{1992A&A...253..173C}. 
The expression of $D_{\rm eff}$ is given by
\begin{equation}
 D_{\rm eff}=\frac{|r U(r)|^2}{30D_{\rm h}}.
\end{equation}
For the vertical shear diffusion coefficient we use the expression of \citet{1997A&A...317..749T} which accounts for the effects of horizontal turbulence:
\begin{equation}
 D_{\rm{shear}} = \frac{(K +D_{\rm h})}{\left[\frac{\varphi}{\delta}\nabla_\mu(1+\frac{K}{D_{\rm h}})+(\nabla_{\rm ad}-\nabla_{\rm rad})\right]} \frac{\alpha H_{\rm p}}{g\delta}\left(0.8836\Omega\frac{d\ln\Omega}{d\ln r}\right)^2,
\label{eq:dshear}
\end{equation}
where $K$ is the thermal diffusivity, the diffusion coefficient $D_{\rm h}$ describes horizontal turbulent transport. As a standard, we used the $D_{\rm h}$ derived by \citet{2003A&A...399..263M}. 

\subsection{Nuclear reaction network}
The nuclear reaction network used previously in the Geneva stellar evolution code included a limited number of isotopes and reaction rates. This prevented the investigation of the evolution of different isotopes as for example the light isotopes lithium, beryllium and boron. We therefore implemented into GENEC the Basel reaction network developed originally by \mbox{F.-K.} Thielemann, which is more flexible in terms of choice of nuclei followed and corresponding reaction rates. This reaction network was previously used in a wide range of astrophysical nucleosynthesis calculations, e.g. \citet{1985ApJ...295..604T}, \citet{2006PhRvL..96n2502F}, etc. The equations describing the abundance changes and the method how these equations are solved are described in \citet{1999JCoAM.109..321H}. The reaction rates are used in their analytical form in the so-called {\sc Reaclib}-format \citep[see][]{2000ADNDT..75....1R}. 

The two reactions which determine the burning timescales of H- and He-burning, $^{14}$N$(p,\gamma)^{15}$O and $3\alpha$ were taken from \citet{2005EPJA...25..455I} and \citet{2005Natur.433..136F} respectively. The analytical fits of these two rates were provided by the JINA reaclib website (groups.nscl.msu.edu/jina/reaclib/db). The rates of the $(p,\alpha)$ and $(p,\gamma)$ reactions involved in the different CNO-cycles were taken from NACRE \citep{1999NuPhA.656....3A}. Also all the $(p,\alpha)$ and $(p,\gamma)$ reactions on the isotopes of lithium, beryllium and boron which are responsible for their destruction come from this source. The involved $\beta^+$-decays are experimental rates which can also be found in the JINA reaclib database (under label ``bet+''). 

We included 43 isotopes from hydrogen up to silicon in the reaction network. These isotopes are listed in Table~\ref{tab:iniabundances}. $^8$Be is included implicitly, i.e. assumed to decay instantaneously into two $\alpha$-particles. 

\begin{table}
\centering
\caption{Isotopes considered in the reaction network and their initial abundance in mass fractions.}
\begin{tabular}{lr|lr}
\hline\hline\\
Isotope & Mass Fraction & Isotope & Mass Fraction\\
\hline\\
n 	&	0.000E-00 & $^{17}$O&	2.266E-06 \\
p 	&	7.200E-01 & $^{18}$O&	1.290E-05 \\
D 	&	1.397E-05 & $^{17}$F&	0.000E-00\\
$^3$He	&	4.415E-05 & $^{18}$F&	0.000E-00\\
$^4$He	&	2.660E-01 & $^{19}$F&	5.407E-07 \\
$^6$Li	&	4.004E-12 & $^{20}$Ne&	1.877E-03\\
$^7$Li	&	5.689E-11 & $^{21}$Ne&	4.724E-06\\
$^7$Be	&	0.000E-00 & $^{22}$Ne&	1.518E-04\\
$^8$Be	&	0.000E-00 & $^{21}$Na&	0.000E-00\\
$^9$Be	&	1.692E-10 & $^{22}$Na&	0.000E-00\\
$^{8}$B &	0.000E-00 & $^{23}$Na&	2.666E-05\\
$^{10}$B&	7.786E-10 & $^{24}$Mg&	5.035E-04\\
$^{11}$B&	3.465E-09 & $^{25}$Mg&	6.641E-05\\
$^{11}$C&	0.000E-00 & $^{26}$Mg&	7.599E-05\\
$^{12}$C&	2.283E-03 & $^{25}$Al&	0.000E-00\\
$^{13}$C&	2.771E-05 & $^{26}$Al&	0.000E-00\\
$^{14}$C&	0.000E-00 & $^{27}$Al&	4.961E-05\\
$^{13}$N&	0.000E-00 & $^{27}$Si&	0.000E-00\\
$^{14}$N&	6.588E-04 & $^{28}$Si&	6.550E-04\\
$^{15}$N&	2.595E-06 & $^{29}$Si&	3.445E-05\\
$^{15}$O&	0.000E-00 & $^{30}$Si&	2.349E-05\\
$^{16}$O&	5.718E-03 &\\
\hline
\end{tabular}
\label{tab:iniabundances}
\end{table}
\begin{figure*}
  \includegraphics[width=0.33\textwidth]{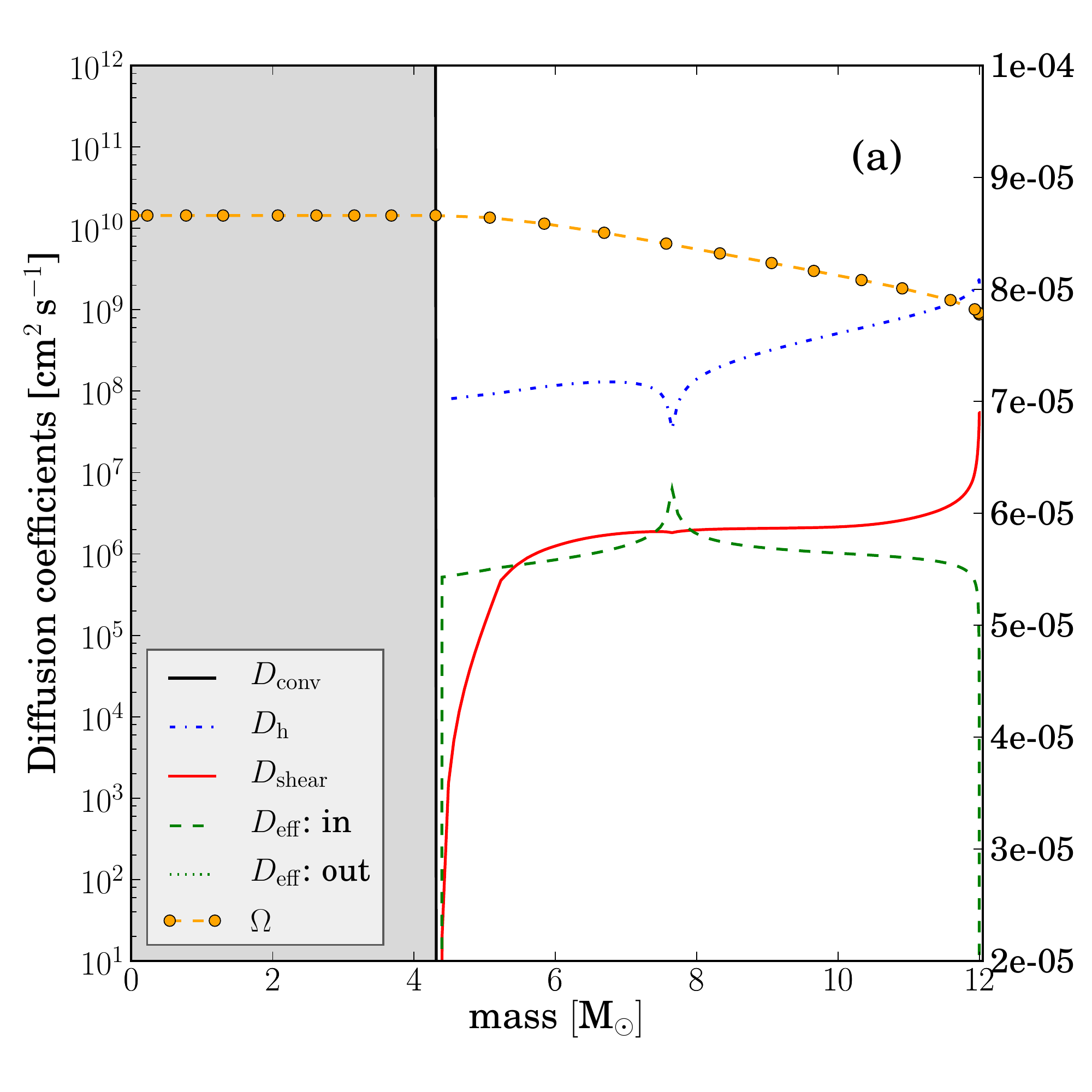} 
  \includegraphics[width=0.33\textwidth]{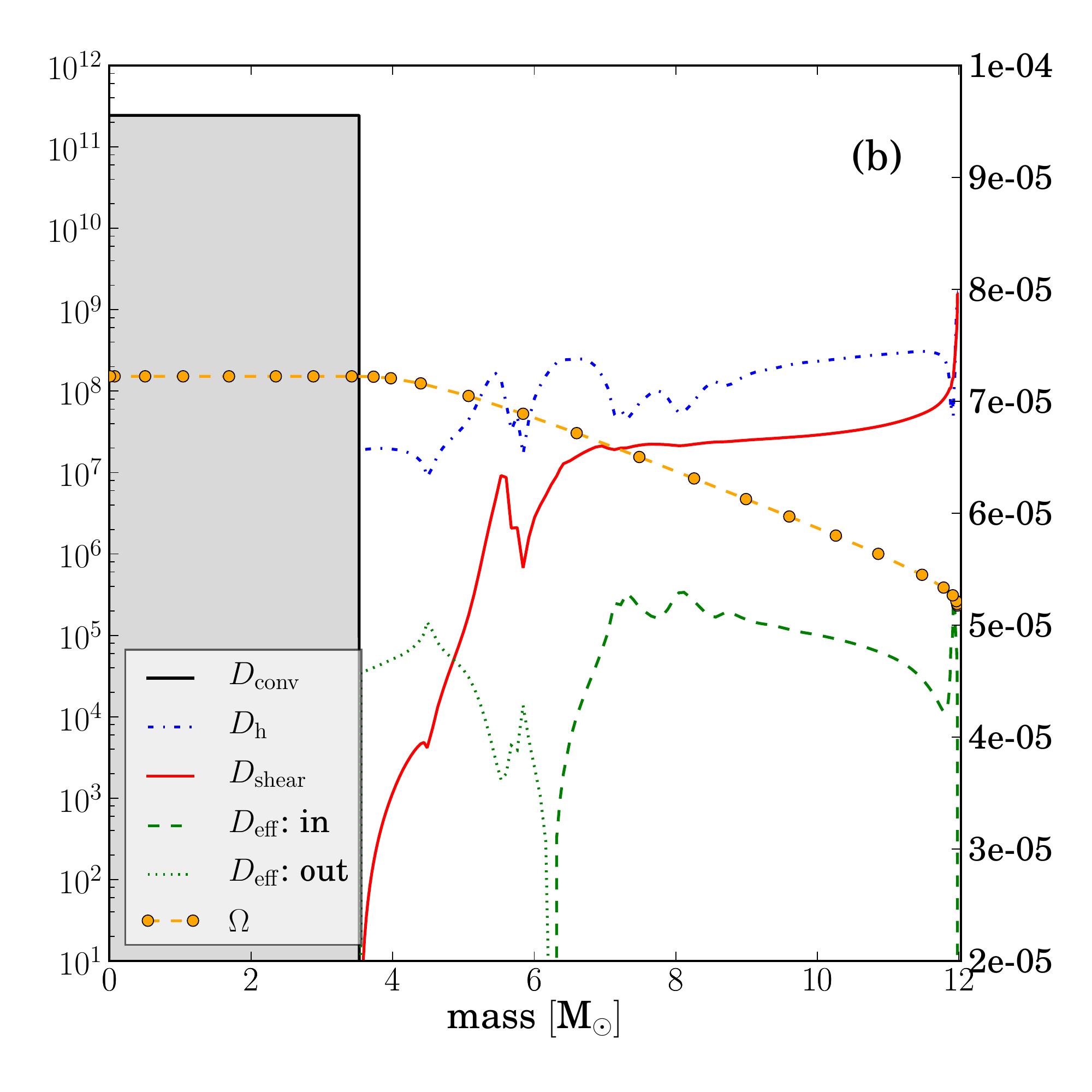} 
  \includegraphics[width=0.33\textwidth]{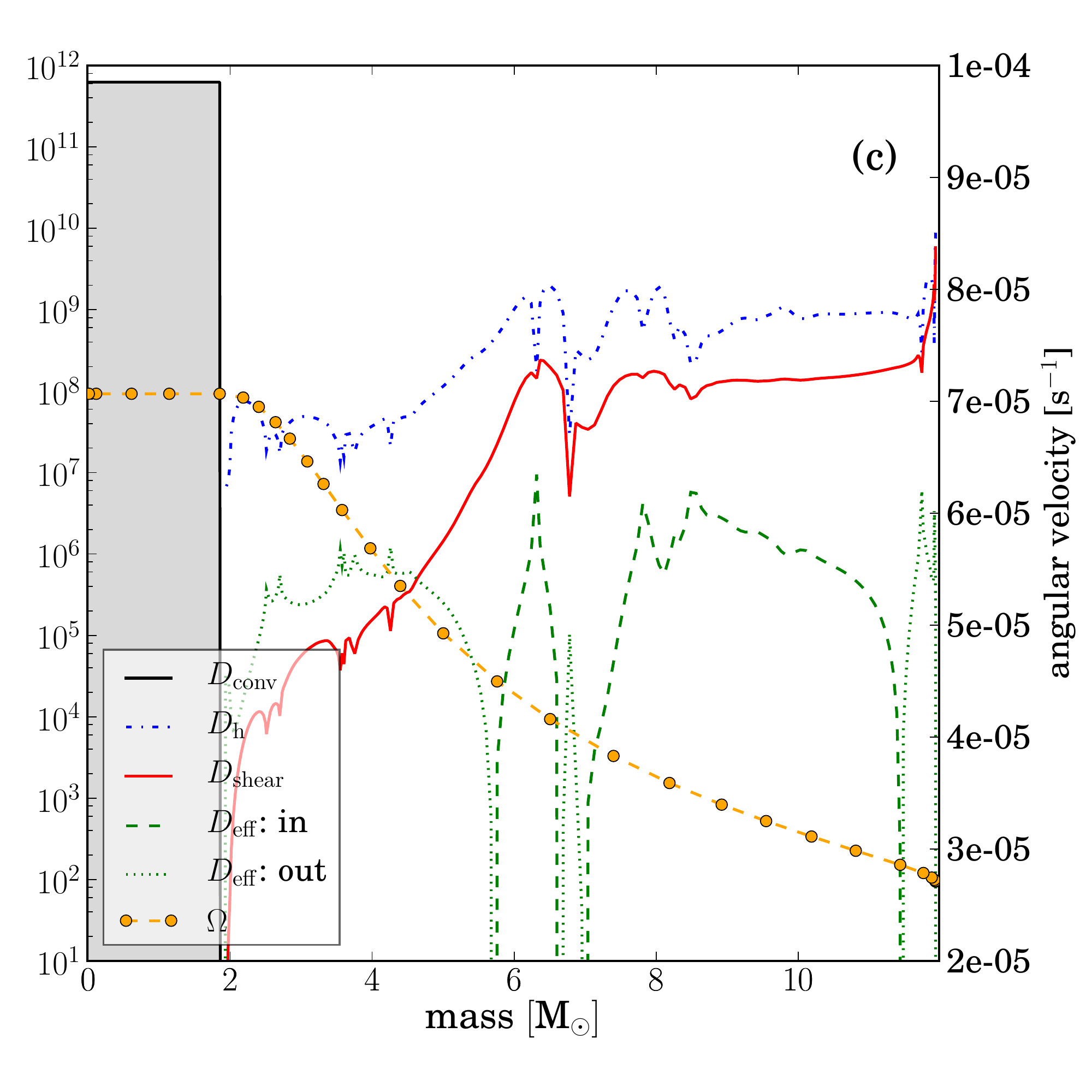}
  \caption{Diffusion coefficients and angular velocity versus mass for a 12M$_\odot$ star with $\upsilon_{\rm ini}/\upsilon_{\rm crit}=0.4$ at the start of hydrogen burning (a) when central hydrogen mass fraction is 0.48 (b) and $5\cdot10^{-3}$ (c). The area below the convective diffusion coefficient $D_{\rm conv}$ is grey shaded to depict the convective core. The diffusion coefficient describing meridional  circulation $D_{\rm eff}$ is plotted with two different lines to distinguish between $U(r)<0$ (green dotted line) and $U(r)>0$ (green dashed line), i.e. the zones transporting angular momentum outwards and inwards respectively.}
  \label{fig:diffcoef} 
\end{figure*}

\subsection{Model grid}
The most recent abundance determinations of boron were based on the B III 2065.8 \r{A} line strength. \citet{2002ApJ...565..571V} found a plateau of the line strength for $T_{\rm eff}$ between 18'000 and 29'000~K. Such surface temperatures are reached by MS stars between about 9 to 15~M$_\odot$. Therefore we chose models with initial masses of 9, 12 and 15~M$_\odot$ to investigate the effect of mixing. For each mass, models with different rotational velocities were calculated. All models were started from the zero age main sequence (ZAMS) and followed up to the end of He-core burning. The time averaged equatorial velocities on the MS phase $\langle\upsilon_{\rm eq}\rangle$ are between 0 to 350~km~s$^{-1}$ (see Table~\ref{tab:models}). The models with rotation were started at ZAMS with a flat angular velocity profile.

Most of our models were started with a solar like composition, since the observations of boron in B-type stars have been made for stars in the solar neighbourhood. As initial composition we chose X=0.72, Y=0.266 and Z=0.014 with the elemental composition from \citet{2005ASPC..336...25A} but the neon abundance from \citet{2006ApJ...647L.143C} and the isotopic percentage from \citet{2003ApJ...591.1220L}. Three models were computed with X=0.7, Y=0.28, Z=0.02 for comparison with models of \citet{2000ApJ...544.1016H}. To investigate the influence of lower metallicity on the surface mixing an additional 12~M$_\odot$ model was calculated with X=0.736, Y=0.257, Z=0.007.

In Table~\ref{tab:models} some parameters of the performed models are presented. In the first column the initial stellar mass M$_{\rm ini}$ is listed and thereafter, equatorial velocity over critical velocity $\upsilon_{\rm ini}/\upsilon_{\rm crit}$, initial angular momentum $J_{\rm ini}$, MS life time $\tau_{\rm H}$,  equatorial velocity $\upsilon_{\rm eq}$, and the surface mass fractions of p, $^{3}$He, $^{4}$He, $^{10}$B, $^{11}$B,$^{12}$C, $^{13}$C, and $^{14}$N at different times of the evolution. 

\begin{table*}[ht]
\caption{Model parameters} 
{\tiny
\begin{tabular}{ccccr cccccccccc}  
\hline\hline \\[-3pt]
M$_{\rm ini}$ 	& $\frac{\upsilon_{\rm ini}}{\upsilon_{\rm crit}}$   &  $J_{\rm ini}$	&  $\tau_{\rm H}$ &$\upsilon_{\rm eq}$	& p & $^{3}$He & $^{4}$He & $^{10}$B & $^{11}$B &   $^{12}$C &  $^{13}$C & $^{14}$N \\
\scriptsize{$[$M$_\odot]$}	&			& \scriptsize{$[10^{52}$ erg s$]$}& \scriptsize{$[10^7$ yr$]$}	& \scriptsize{$[$km~s$^{-1}]$} & \multicolumn{8}{c}{Mass fraction}\\[3pt]
\hline\\[-5pt]
9			& 0.0		& 0.00	& 2.538 &  0 (0)$^a$	& 0.720 & 6.51E-05 & 0.266 & 7.79E-10 & 3.46E-09 & 2.28E-03 & 2.77E-05 & 6.59E-04 \\
			&		&	&	&  0$^b$	& 0.720 & 6.51E-05 & 0.266 & 7.79E-10 & 3.46E-09 & 2.28E-03 & 2.77E-05 & 6.59E-04 \\
			&		&	&	&  0$^c$	& 0.703 & 4.40E-05 & 0.283 & 3.84E-11 & 2.66E-10 & 1.36E-03 & 7.01E-05 & 2.18E-03 \\[2pt]
9			& 0.2		& 0.35	& 2.632	& 88 (115)$^a$	& 0.720 & 6.00E-05 & 0.266 & 5.85E-11 & 5.25E-10 & 2.13E-03 & 5.97E-05 & 8.09E-04 \\
			&		&	&	& 76$^b$	& 0.720 & 4.96E-05 & 0.266 & 1.14E-11 & 1.38E-10 & 1.76E-03 & 1.03E-04 & 1.20E-03 \\
			&		&	&	& 1$^c$		& 0.696 & 3.07E-05 & 0.290 & 1.05E-12 & 1.53E-11 & 1.05E-03 & 1.11E-04 & 2.59E-03 \\[2pt]
9			& 0.4		& 0.67	& 2.663 & 182 (231)$^a$	& 0.720 & 3.37E-05 & 0.266 & 2.83E-13 & 7.42E-12 & 1.31E-03 & 1.55E-04 & 1.67E-03 \\
			&		&	&	& 163$^b$	& 0.719 & 2.63E-05 & 0.267 & 1.04E-14 & 4.68E-13 & 9.48E-04 & 1.56E-04 & 2.15E-03 \\
			&		&	&	& 1$^c$		& 0.692 & 1.97E-05 & 0.294 & 9.25E-16 & 5.09E-14 & 6.71E-04 & 1.28E-04 & 3.03E-03 \\[2pt]
9			& 0.6		& 0.95	& 2.692	& 275 (348)$^a$	& 0.719 & 2.22E-05 & 0.267 & 1.50E-15 & 1.01E-13 & 8.25E-04 & 1.53E-04 & 2.28E-03 \\
			&		&	&	& 256$^b$	& 0.717 & 1.84E-05 & 0.269 & 1.07E-17 & 1.57E-15 & 5.49E-04 & 1.28E-04 & 2.74E-03 \\
			&		&	&	& 2$^c$		& 0.687 & 1.47E-05 & 0.298 & 6.87E-19 & 1.48E-16 & 4.04E-04 & 1.01E-04 & 3.43E-03 \\[2pt]
\hline\\[-6pt]
12			& 0.0		& 0.00	& 1.605	& 0 (0)$^a$	& 0.720 & 6.51E-05 & 0.266 & 7.79E-10 & 3.46E-09 & 2.28E-03 & 2.77E-05 & 6.59E-04 \\
			&		&	&	& 0$^b$		& 0.720 & 6.51E-05 & 0.266 & 7.79E-10 & 3.46E-09 & 2.28E-03 & 2.77E-05 & 6.59E-04 \\
			&		&	&	& 0$^c$		& 0.684 & 3.77E-05 & 0.302 & 2.30E-11 & 2.35E-10 & 1.33E-03 & 7.15E-05 & 2.42E-03 \\[2pt]
12			& 0.1		& 0.31	& 1.619	& 44 (62)$^a$	& 0.720 & 6.51E-05 & 0.266 & 3.05E-10 & 1.89E-09 & 2.28E-03 & 2.92E-05 & 6.60E-04 \\
			&		&	&	& 36$^b$	& 0.720 & 6.09E-05 & 0.266 & 9.37E-11 & 7.25E-10 & 2.19E-03 & 5.21E-05 & 7.43E-04 \\
			&		&	&	& 0$^c$		& 0.680 & 3.25E-05 & 0.306 & 7.51E-12 & 7.06E-11 & 1.23E-03 & 9.24E-05 & 2.57E-03 \\[2pt]
12			& 0.2		& 0.62	& 1.638	& 94 (123)$^a$	& 0.720 & 5.33E-05 & 0.266 & 2.37E-11 & 2.56E-10 & 2.02E-03 & 8.10E-05 & 9.18E-04 \\
			&		&	&	& 79$^b$	& 0.720 & 4.02E-05 & 0.266 & 2.94E-12 & 4.49E-11 & 1.61E-03 & 1.26E-04 & 1.36E-03 \\
			&		&	&	& 1$^c$		& 0.678 & 2.49E-05 & 0.308 & 2.57E-13 & 4.71E-12 & 1.02E-03 & 1.17E-04 & 2.81E-03 \\[2pt]
12			& 0.3		& 0.91	& 1.663	& 144 (184)$^a$	& 0.720 & 3.63E-05 & 0.266 & 9.93E-13 & 1.98E-11 & 1.56E-03 & 1.47E-04 & 1.40E-03 \\
			&		&	&	& 125$^b$	& 0.719 & 2.67E-05 & 0.267 & 3.92E-14 & 1.31E-12 & 1.19E-03 & 1.68E-04 & 1.85E-03 \\
			&		&	&	& 1$^c$		& 0.677 & 1.88E-05 & 0.309 & 3.32E-15 & 1.34E-13 & 8.39E-04 & 1.38E-04 & 2.98E-03 \\[2pt]
12			& 0.4		& 1.20	& 1.669	& 194 (246)$^a$	& 0.719 & 2.55E-05 & 0.266 & 2.50E-14 & 1.03E-12 & 1.20E-03 & 1.74E-04 & 1.80E-03 \\
			&		&	&	& 173$^b$	& 0.718 & 1.95E-05 & 0.268 & 2.97E-16 & 2.43E-14 & 9.05E-04 & 1.73E-04 & 2.23E-03 \\
			&		&	&	& 1$^c$		& 0.676 & 1.47E-05 & 0.310 & 2.38E-17 & 2.38E-15 & 6.67E-04 & 1.39E-04 & 3.19E-03 \\[2pt]
12			& 0.5		& 1.46	& 1.676	& 245 (309)$^a$	& 0.719 & 1.96E-05 & 0.267 & 7.29E-16 & 5.69E-14 & 9.57E-04 & 1.76E-04 & 2.12E-03 \\
			&		&	&	& 224$^b$	& 0.716 & 1.56E-05 & 0.269 & 2.42E-18 & 4.73E-16 & 6.98E-04 & 1.60E-04 & 2.55E-03 \\
			&		&	&	& 1$^c$		& 0.676 & 1.24E-05 & 0.310 & 8.87E-20 & 4.36E-17 & 5.33E-04 & 1.29E-04 & 3.35E-03 \\[2pt]
12			& 0.6		& 1.68	& 1.681	& 295 (371)$^a$	& 0.719 & 1.62E-05 & 0.267 & 3.51E-17 & 4.47E-15 & 7.85E-04 & 1.68E-04 & 2.36E-03 \\
			&		&	&	& 276$^b$	& 0.715 & 1.34E-05 & 0.271 & 2.72E-22 & 1.39E-17 & 5.51E-04 & 1.42E-04 & 2.80E-03 \\
			&		&	&	& 1$^c$		& 0.692 & 1.18E-05 & 0.293 & 4.15E-24 & 1.21E-18 & 4.65E-04 & 1.23E-04 & 3.24E-03 \\[2pt]
12			& 0.7		& 1.87	& 1.698	& 344 (431)$^a$	& 0.718 & 1.41E-05 & 0.267 & 3.19E-18 & 5.71E-16 & 6.81E-04 & 1.63E-04 & 2.50E-03 \\
			&		&	&	& 337$^b$	& 0.713 & 1.20E-05 & 0.273 & 7.73E-27 & 5.07E-20 & 4.54E-04 & 1.27E-04 & 2.98E-03 \\
			&		&	&	& 2$^c$		& 0.660 & 9.22E-06 & 0.326 & 8.21E-29 & 5.38E-22 & 3.39E-04 & 9.72E-05 & 3.79E-03 \\[2pt]
12			& 0.4		& 1.20	& 1.667	& 195 (248)$^{a,d}$ & 0.735 & 1.95E-05 & 0.258 & 3.48E-16 & 3.11E-14 & 4.17E-04 & 7.33E-05 & 1.16E-03 \\
			&		&	&	& 173$^{b,d}$	& 0.733 & 1.52E-05 & 0.260 & 7.39E-19 & 2.67E-16 & 2.83E-04 & 6.38E-05 & 1.38E-03 \\
			&		&	&	& 1$^{c,d}$	& 0.727 & 1.41E-05 & 0.266 & 1.32E-20 & 2.91E-17 & 2.47E-04 & 5.89E-05 & 1.50E-03 \\[2pt]
\hline\\[-6pt]
15			& 0.0		& 0.00	& 1.128	& 0 (0)$^a$	& 0.720 & 6.51E-05 & 0.266 & 7.79E-10 & 3.46E-09 & 2.28E-03 & 2.77E-05 & 6.59E-04 \\
			&		&	&	& 0$^b$		& 0.720 & 6.51E-05 & 0.266 & 7.78E-10 & 3.46E-09 & 2.28E-03 & 2.77E-05 & 6.59E-04 \\
			&		&	&	& 0$^e$		& 0.680 & 3.14E-05 & 0.305 & 2.99E-16 & 6.92E-11 & 1.17E-03 & 6.73E-05 & 2.70E-03 \\[2pt]
15			& 0.2		& 0.96	& 1.197	& 97 (129)$^a$	& 0.720 & 4.88E-05 & 0.266 & 8.58E-12 & 1.19E-10 & 1.95E-03 & 9.93E-05 & 9.84E-04 \\
			&		&	&	& 80$^b$	& 0.719 & 3.56E-05 & 0.267 & 6.82E-13 & 1.48E-11 & 1.56E-03 & 1.45E-04 & 1.43E-03 \\
			&		&	&	& 0$^e$		& 0.661 & 2.01E-05 & 0.325 & 4.68E-14 & 1.24E-12 & 9.18E-04 & 1.19E-04 & 3.17E-03 \\[2pt]
15			& 0.4		& 1.85	& 1.209	& 203 (257)$^a$	& 0.719 & 2.23E-05 & 0.267 & 3.59E-15 & 2.15E-13 & 1.22E-03 & 1.96E-04 & 1.80E-03 \\
			&		&	&	& 179$^b$	& 0.716 & 1.69E-05 & 0.270 & 2.07E-17 & 2.68E-15 & 9.54E-04 & 1.94E-04 & 2.22E-03 \\
			&		&	&	& 0$^e$		& 0.650 & 1.21E-05 & 0.336 & 1.26E-18 & 2.13E-16 & 6.78E-04 & 1.47E-04 & 3.42E-03 \\[2pt]
15			& 0.6		& 2.61	& 1.224	& 308 (387)$^a$	& 0.718 & 1.42E-05 & 0.268 & 1.72E-18 & 3.86E-16 & 8.55E-04 & 1.93E-04 & 2.29E-03 \\
			&		&	&	& 286$^b$	& 0.712 & 1.14E-05 & 0.274 & 6.07E-28 & 4.13E-20 & 6.24E-04 & 1.65E-04 & 2.76E-03 \\
			&		&	&	& 1$^e$		& 0.642 & 8.30E-06 & 0.344 & 4.15E-30 & 2.82E-22 & 4.49E-04 & 1.22E-04 & 3.79E-03 \\[2pt]
\hline\\[-5pt]
\end{tabular}}
\\ $^a$ The surface velocity and mass fractions when the central hydrogen abundance X$_c\approx0.36$ is given on the first line for each model. The velocity values in brackets are the values at ZAMS.
\\ $^b$ The surface velocity and mass fractions at the end of hydrogen burning (X$_c\approx10^{-5}$) is given on the second line for each model. The velocity was not taken exactly when the hydrogen was depleted but when it reached its minimum, i.e., a little bit earlier. 
\\ $^c$ The third line for the 9 and 12~M$_\odot$ models corresponds to the surface values during the RGB phase.
\\ $^d$ model with lower initial metallicity Z$=0.5\cdot$Z$_\odot$. 
\\ $^e$ The third line for the 15~M$_\odot$ models corresponds to the surface values at the end of He-burning.
\label{tab:models}
\end{table*}

\section{Models}
\label{sec:results}
\subsection{Rotation and mixing}
In Fig.~\ref{fig:diffcoef} the diffusion coefficients in a 12~M$_\odot$ model with intermediate rotation velocity ($\langle \upsilon_{\rm eq}\rangle\approx$200~km~s$^{-1}$) are plotted for three different times on the MS. When meridional circulation currents descend at the equator and ascend at the pole, i.e. when it transports momentum towards the centre, $D_{\rm eff}$ is drawn as (green) dashed line and for the opposite circulation direction as dotted line.

All the models begin their evolution on the ZAMS with a flat angular velocity profile ($\Omega=$constant). The profile of $\Omega$ converges rapidly towards an ``equilibrium'' profile where the advection of angular momentum towards the inner layers is compensated by the  diffusion of angular momentum towards the outer layers \citep[see][]{1999A&A...341..181D,2000A&A...361..101M}. The slow expansion of the stellar outer layers during the MS phase, the core contraction and the effects of meridional currents and shear diffusion, lead to a continuous and slow change of $\Omega$.

Very early in the evolution, a situation with two cells of meridional currents sets in: an inner shell which brings angular momentum towards the surface and an outer cell which transport angular momentum inwards (see Fig.~\ref{fig:diffcoef}b). Close to the end of the MS even more meridional current cells appear (see Fig.~\ref{fig:diffcoef}c). Except for a short while at the very beginning of the evolution, the transport of the chemical species is mainly due to $D_{\rm eff}$ near the convective core and to $D_{\rm shear}$ in the outer part of the radiative zone. $D_{\rm shear}$ is thus the key parameter responsible for boron depletion at the surface. The nitrogen enhancement at the surface is due to the effects of both $D_{\rm eff}$ and $D_{\rm shear}$ since nitrogen is enhanced in the convective core and thus must be transported through the whole radiative envelope.

\begin{figure*}
  \includegraphics[width=0.33\textwidth]{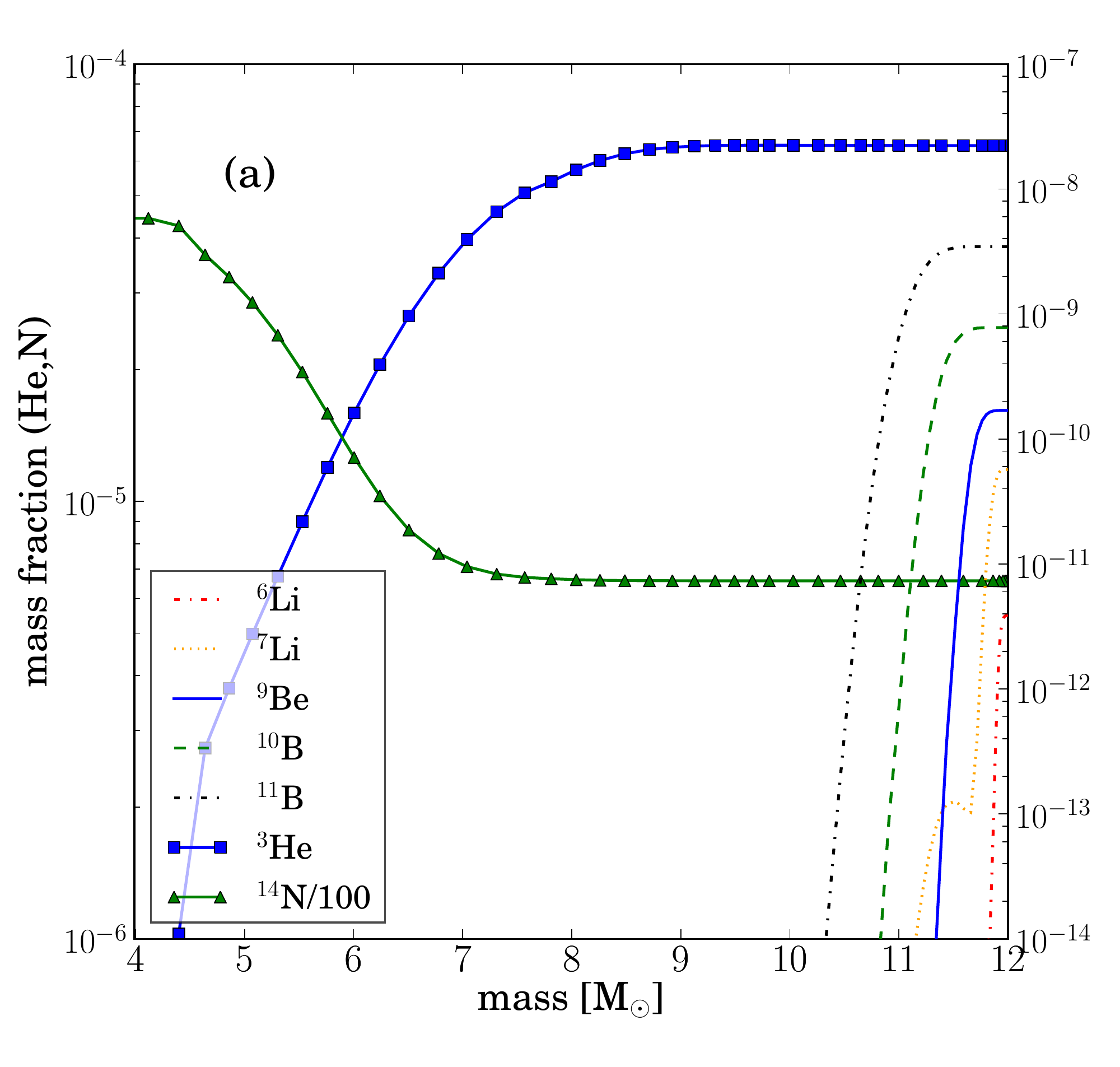} 
  \includegraphics[width=0.33\textwidth]{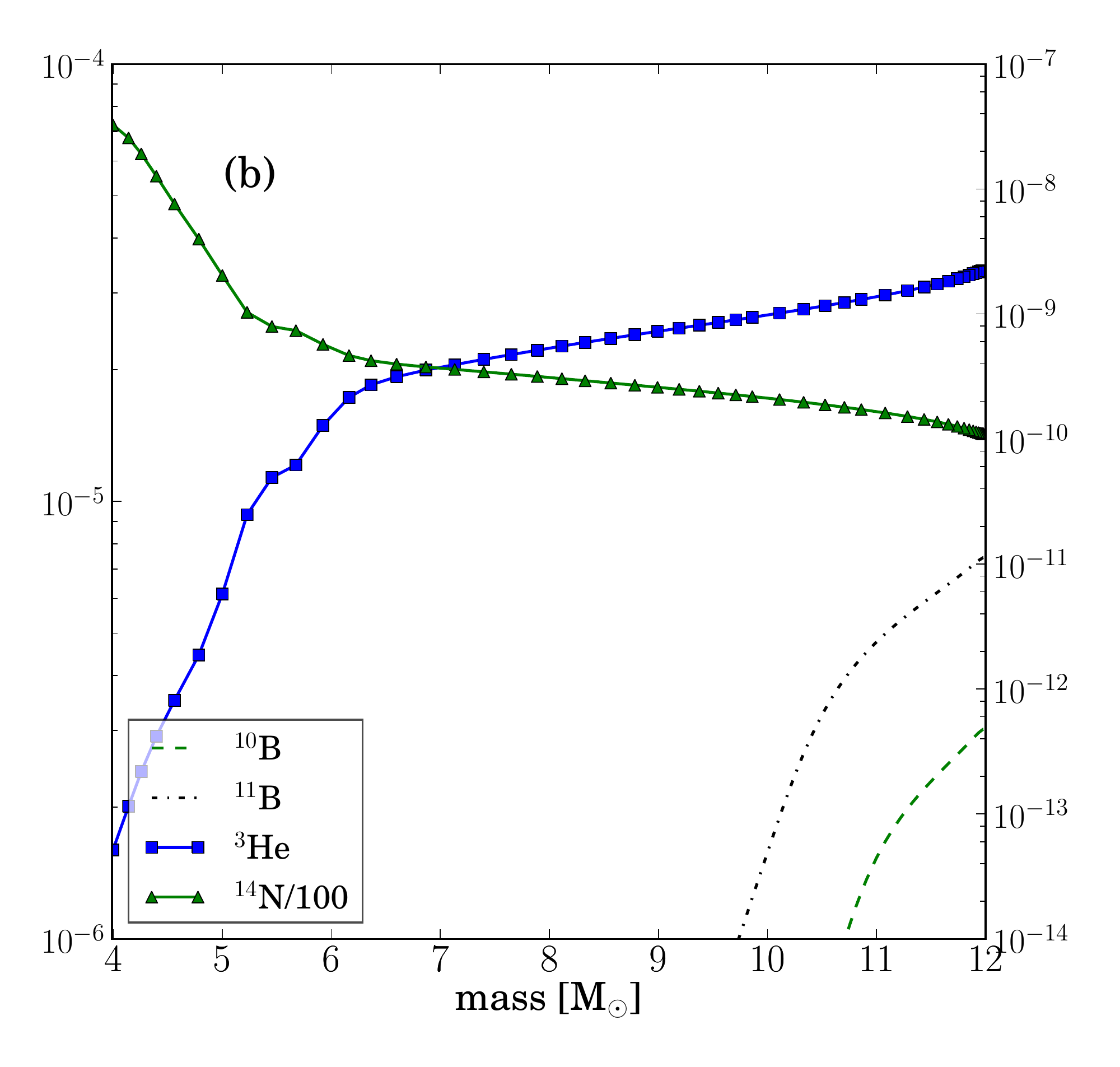} 
  \includegraphics[width=0.33\textwidth]{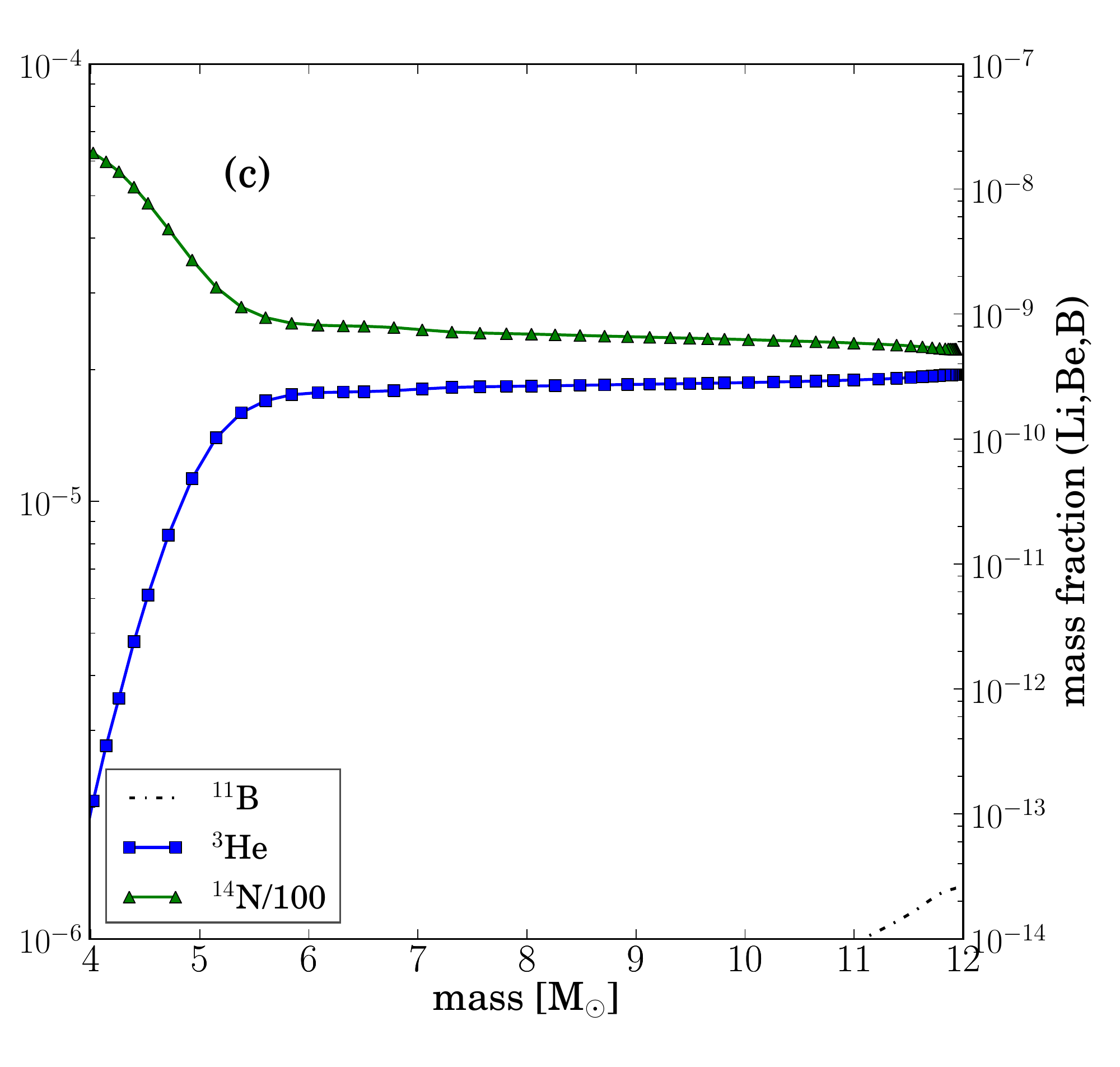}
  \caption{Mass fractions of light isotopes versus mass for a 12~M$_\odot$ star with $\upsilon_{\rm ini}/\upsilon_{\rm crit}=0.4$ at the start of hydrogen burning (a) when central hydrogen mass fraction is 0.48 (b) and $5\cdot10^{-3}$ (c). In this model the mixing is strong enough to deplete boron almost completely at the end of the MS, lithium and beryllium are depleted at much earlier times.}
  \label{fig:lightelemabund} 
\end{figure*}
  
\subsection{Evolution of surface composition}
In Fig.~\ref{fig:lightelemabund} the variations with the Lagrangian mass coordinate of various elements are shown. At the beginning of the evolution, on the ZAMS, there is a thin surface layer containing Li, Be and B. The mass of this surface layer is between 0.3 and 2~M$_\odot$ depending on the isotope and the model considered. We see that the isotopes of Be and Li disappear very rapidly from the surface (and therefore from the whole star!). The boron isotopes are also rapidly depleted at the surface although on a longer timescale than Li. Changes of the abundances of $^3$He and $^{14}$N at the surface take still more time.

The physical reasons for these different timescales associated to the changes in surface abundances are the different temperatures required to change the abundances of these elements by nuclear reactions: the Li isotopes are destroyed at about $3\cdot10^6$~K, Be and B isotopes start to be depleted as soon as the temperature reaches about 5 and 6$\cdot10^6$~K respectively, $^3$He and $^{14}$N  still need higher temperatures to be destroyed/synthesized (in case of nitrogen) of the order of $1.4\cdot10^7$ (for $^3$He) and of $1.7\cdot10^7$~K (for nitrogen). In the absence of any mixing in the radiative zones, as would be the case in standard non-rotating models, the surface abundances of these elements would not change. When some mixing processes are at work, as those induced by rotation, one expects changes of the surface abundances, more rapid for those elements whose abundances are changed at low temperatures, near the surface (such as Li, Be, and B), than for those which are depleted/synthesized in deeper layers (such as $^3$He and $^{14}$N). For instance, boron depletion is obtained by transport processes in a much smaller portion of the star than the one required to obtain nitrogen enhancement. Nitrogen indeed needs to be transported through the whole radiative envelope, while boron only needs to be transported through a small part of it. Therefore present models predict the existence of boron depleted stars with no nitrogen enrichments. From Fig.~\ref{fig:lightelemabund}, we see also that the ratio of $^{11}$B/$^{10}$B increases when evolution goes on, because $^{10}$B is destroyed closer to the surface than $^{11}$B but this features is probably not observable since it occurs when both isotopes are already strongly depleted.

\begin{figure*}
 \includegraphics[width=0.485\textwidth]{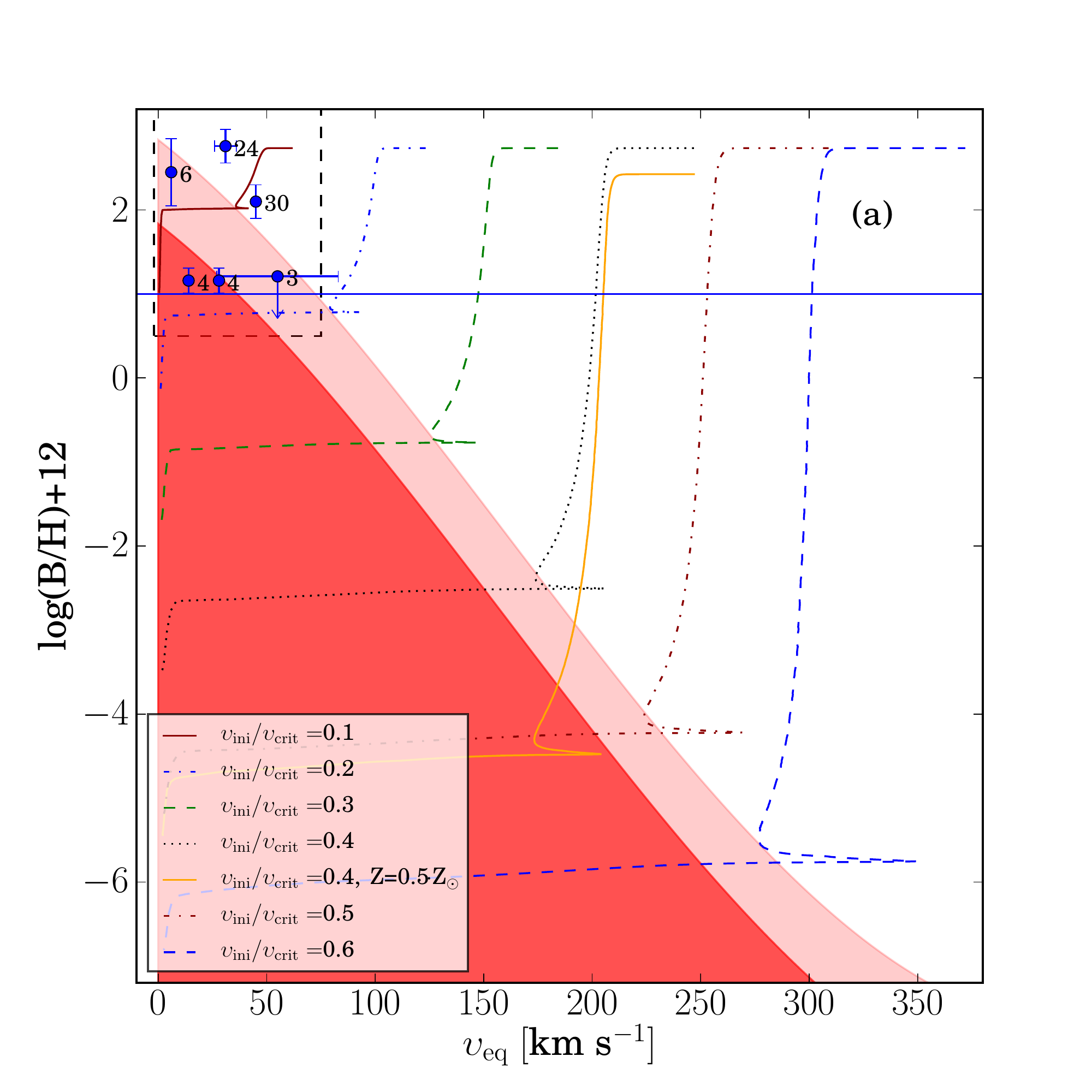}
 \includegraphics[width=0.485\textwidth]{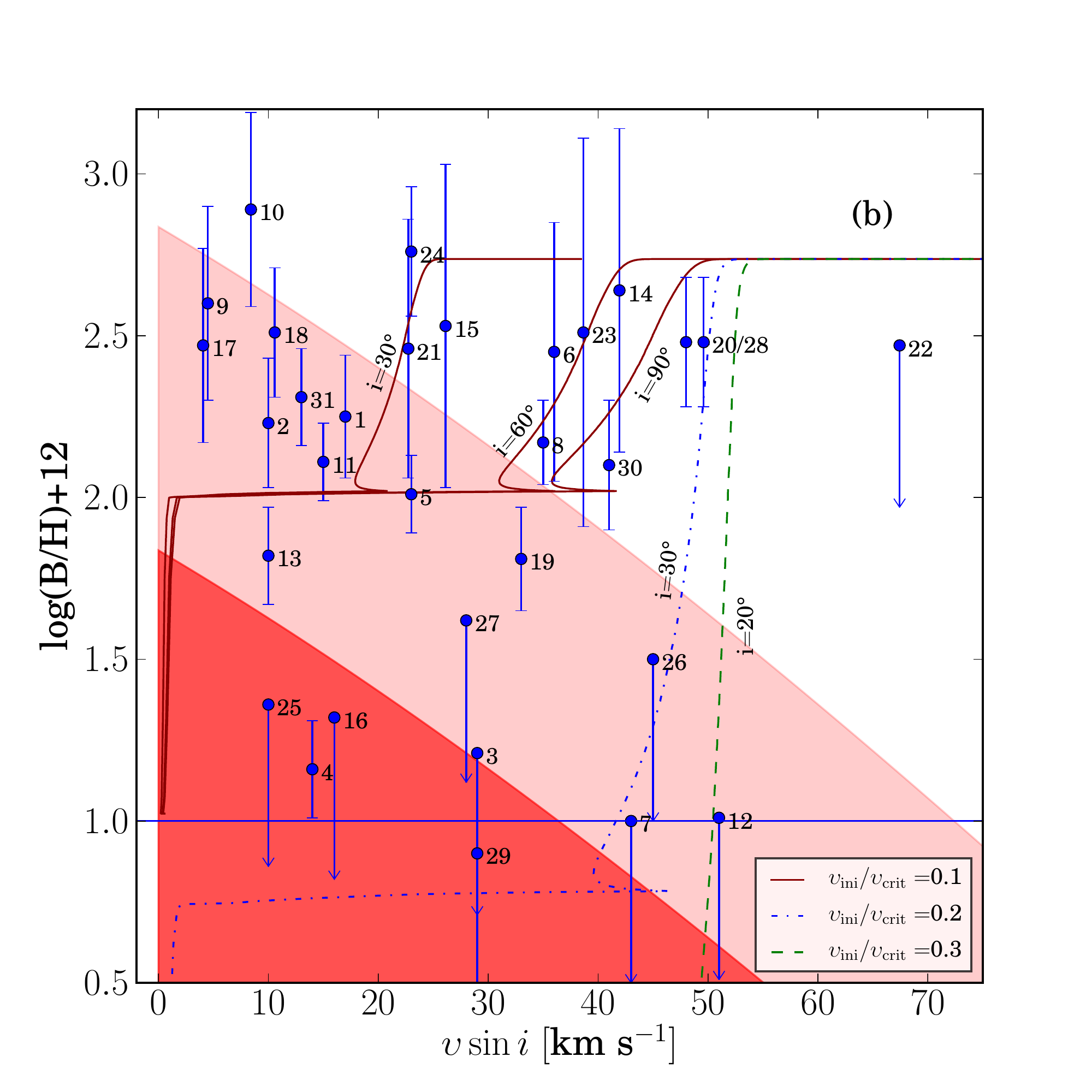}
 \caption{Boron versus equatorial velocity for 12~M$_\odot$ models with different rotation velocities. The stars from the chosen sample with known rotational velocities are shown in (a) and all stars with their $\upsilon\sin i$ in (b). The horizontal line indicates the minimum boron abundance that can still be detected. In (a) the dashed rectangle illustrates the section shown in (b). The red hatched area shows the evolution region for the post MS phases of our models (assuming i=90\textdegree), whereas the light red area could still be on the MS for models with sub-solar Z. The stars are labelled by the number given in the first column of Tables~\ref{tab:obsdata} and \ref{tab:abundobs}. Star 4 has either $\upsilon_{\rm eq}=$14 or 28~km~s$^{-1}$ \citep[see][]{2006ApJ...642..470A}. In (b) the curves depict models with different inclination angles. }
 \label{fig:logBvsveq}
\end{figure*}

Figure~\ref{fig:logBvsveq} shows how boron depletion occurs at the surface when different initial rotation velocities
(and therefore angular momentum content) are considered. In this diagram, evolution proceeds from right to left. Non-rotating models would show non-depleted boron surface abundances during the whole MS phase. Only when the star is at the red supergiant stage the model predicts a lowering of the surface abundance in boron (log$\epsilon$(B)\footnote{log$\epsilon$(X):=log(Y(X)/Y(H))+12 with Y(X) the number abundance of element X} drops down to 1.59). This is due to the dilution of the boron-rich outer layer with deeper boron depleted layers when an external convective zone appears. The evolution in Fig.~\ref{fig:logBvsveq} of rotating models is quite different with depletion of boron already during the MS evolution. We can see a first phase during which the surface velocity decreases, while no changes of the surface boron abundance occur. The time spent during that phase depends on the initial rotation. As a numerical example, this first phase lasts about 6~Myr for the 12~M$_\odot$ model with $\upsilon_{\rm ini}/\upsilon_{\rm{crit}}=0.1$ and about 0.6~Myr for the 12~M$_\odot$ model with $\upsilon_{\rm ini}/\upsilon_{\rm{crit}}=0.6$. During a second phase, the surface abundance decreases. The decrease occurs nearly at constant surface velocity in the case of the  $\upsilon_{\rm ini}/\upsilon_{\rm{crit}}=0.6$ model, indicating that the mixing timescale is very rapid. It occurs on a longer timescale for lower initial rotation rates. Interestingly, we see that stars with a low initial rotation (below about $\upsilon_{\rm ini}/\upsilon_{\rm{crit}}=0.2$) still have observable boron surface abundances (log$\epsilon$(B)$\ge$1) at the end of the MS phase. This is an interesting feature. Indeed boron on the surface of stars in the HR gap would tell us that these stars had a small rotation rates during the previous phases. Another important point is that boron depletion is very sensitive to the metallicity. Stars with sub-solar metallicity are more compact and undergo enhanced mixing. They end up with a stronger boron depletion for the same evolutionary stage on the MS. Our 12~M$_\odot$ model with half solar metallicity reaches a lower boron surface abundance by 2~dex at the end of its MS life.  

Beyond the MS all models develop a convective zone at the surface, in which the remaining boron is diluted but not burned, since the temperature at the bottom of the convective zone is only about $6\cdot 10^5$~K. For the very slow rotators ($\upsilon_{\rm ini}/\upsilon_{\rm{crit}}\le 0.1$, $\langle\upsilon_{\rm eq}\rangle\le$~50km~s$^{-1}$ on the MS) our models predict observable boron abundances (log$\epsilon$(B)$>$1) even in the red supergiant phase.   

\begin{figure}
 \includegraphics[width=0.485\textwidth]{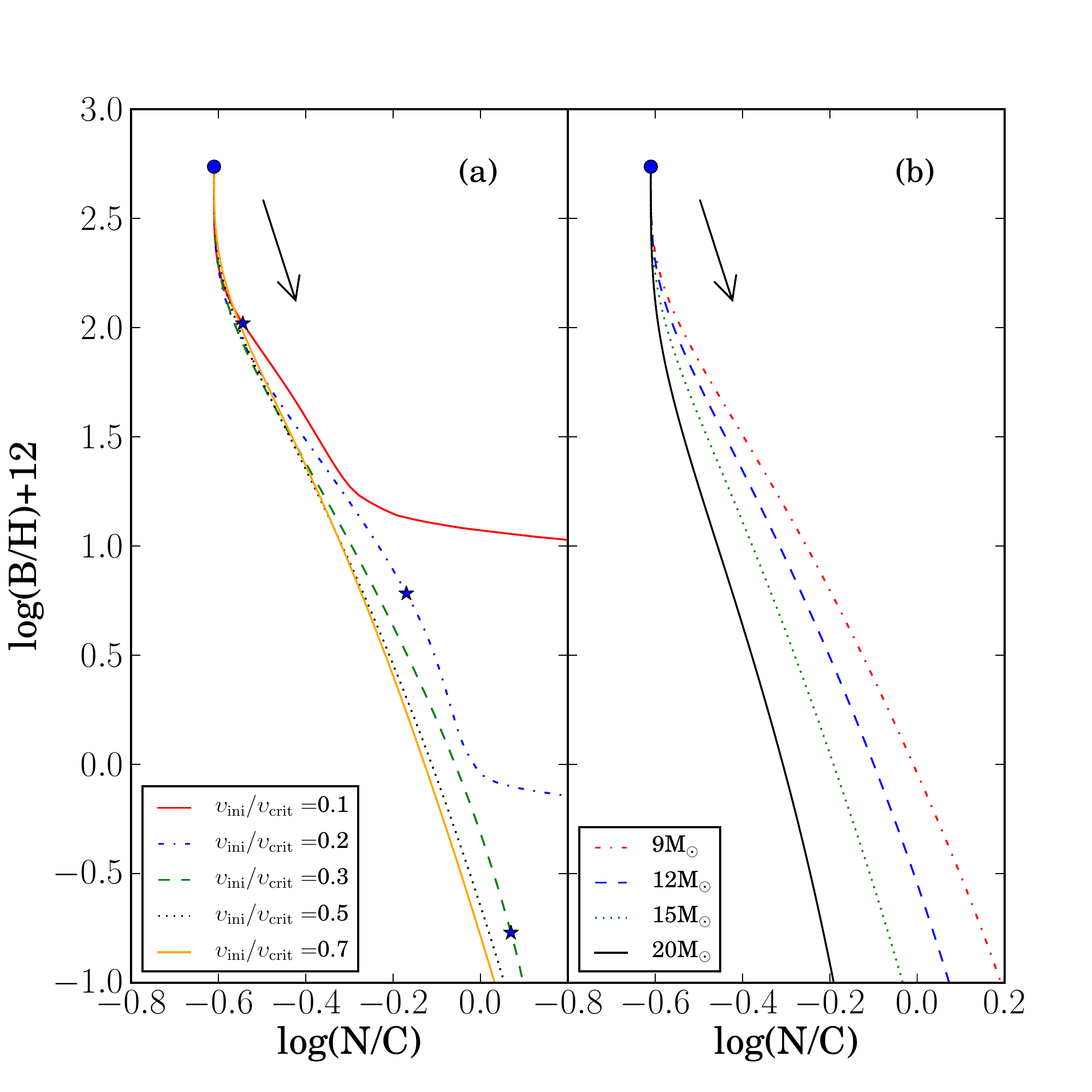}
 \caption{Boron versus log(N/C) for 12~M$_\odot$ models with different rotation velocities (a) and of 9, 12, 15, and 20~M$_\odot$ models with $\upsilon_{\rm ini}/\upsilon_{\rm{crit}}=$0.4 (b). In the left hand plot, the end of the MS of the slow rotators ($\upsilon_{\rm ini}/\upsilon_{\rm{crit}}=$0.1-0.3) is marked by a star symbol. The ZAMS position is indicated by a circle.}
 \label{fig:logBvslogNC_models}
\end{figure}

The evolution of boron and nitrogen abundances at the surface of our stellar models is drawn in Fig.~\ref{fig:logBvslogNC_models}. The 12~M$_\odot$ models with $\upsilon_{\rm ini}/\upsilon_{\rm{crit}}$ between 0.1 and 0.7 follow a similar path (see Fig.~\ref{fig:logBvslogNC_models}a), with the exception that the faster rotators have larger changes in nitrogen and boron by the end of the MS. Thus we see that a change in the initial velocity mainly affects the timescales for the changes of the surface abundances (more rapid with higher rotation rates) but not the correlation much between the abundances of these two elements during the MS phase. 

We also see that similar correlations are found for stars of different initial masses (see Fig.~\ref{fig:logBvslogNC_models}b). We can, however, note that the lower the initial mass, the stronger the nitrogen surface enrichment at a given boron abundance. This is because a given boron abundance is reached after a significantly longer time in the 9~M$_\odot$ stellar model than in the 15~M$_\odot$ one, thus giving more time for changes in nitrogen in the surface layers of the 9~M$_\odot$ model. 

\begin{figure}
 \includegraphics[width=0.485\textwidth]{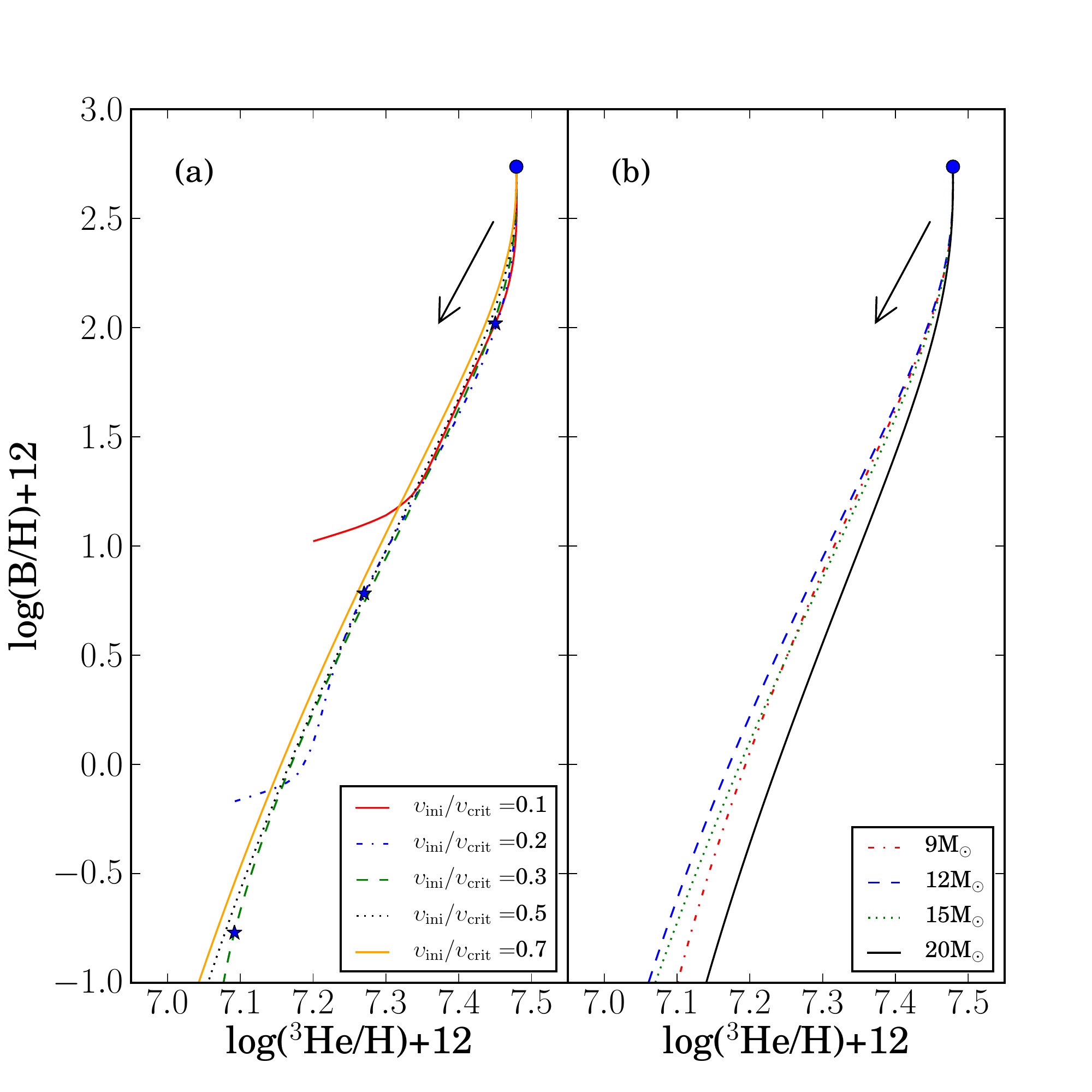}
 \caption{Boron versus log($^3$He/H) for 12~M$_\odot$ models with different rotation velocities (a) and of 9, 12, 15, and 20~M$_\odot$ models with $\upsilon_{\rm ini}/\upsilon_{\rm{crit}}=$0.4 (b). In the left hand plot, the end of the MS of the slow rotators ($\upsilon_{\rm ini}/\upsilon_{\rm{crit}}=$0.1-0.3) is marked by a star symbol. The ZAMS position is indicated by a circle.}
 \label{fig:logBvslogHe3} 
\end{figure}

The depletion of boron is also correlated with changes in other abundances. The case of $^3$He is shown in Fig.~\ref{fig:logBvslogHe3}. The nuclear reactions affecting $^3$He occur at higher temperature than those affecting boron and at a lower temperature than those affecting nitrogen. Thus the changes on the surface of $^3$He occur more rapidly than those of nitrogen but less rapidly than those of boron.

It is interesting to look at possible correlations between surface abundances of boron and $^4$He since the abundances of these two elements can be obtained by spectroscopy for OB-type stars. Helium enrichments have, for instance, been obtained by \citet[][]{2004MNRAS.351..745L}. From a theoretical point of view, one expects that the changes in surface helium abundance take much more time than changes in nitrogen. This comes from the fact that the abundance gradient of helium that builds up at the border of the convective core  is quite shallow with respect to the gradient in the abundance of nitrogen, and the stronger the gradient, the more rapid the diffusion \citep[see e.g. Eq. 3 in ][]{2004A&A...416.1023M}. The gradient of nitrogen is steeper than the one of helium because nitrogen is very rapidly enhanced in the core as a result of the CN cycle, while it takes much longer timescales to increase the central helium abundance. One consequence is that the present models predict that, as long as boron is observable at the surface on the MS, no helium enrichment is predicted.

\begin{figure}
 \includegraphics[width=0.485\textwidth]{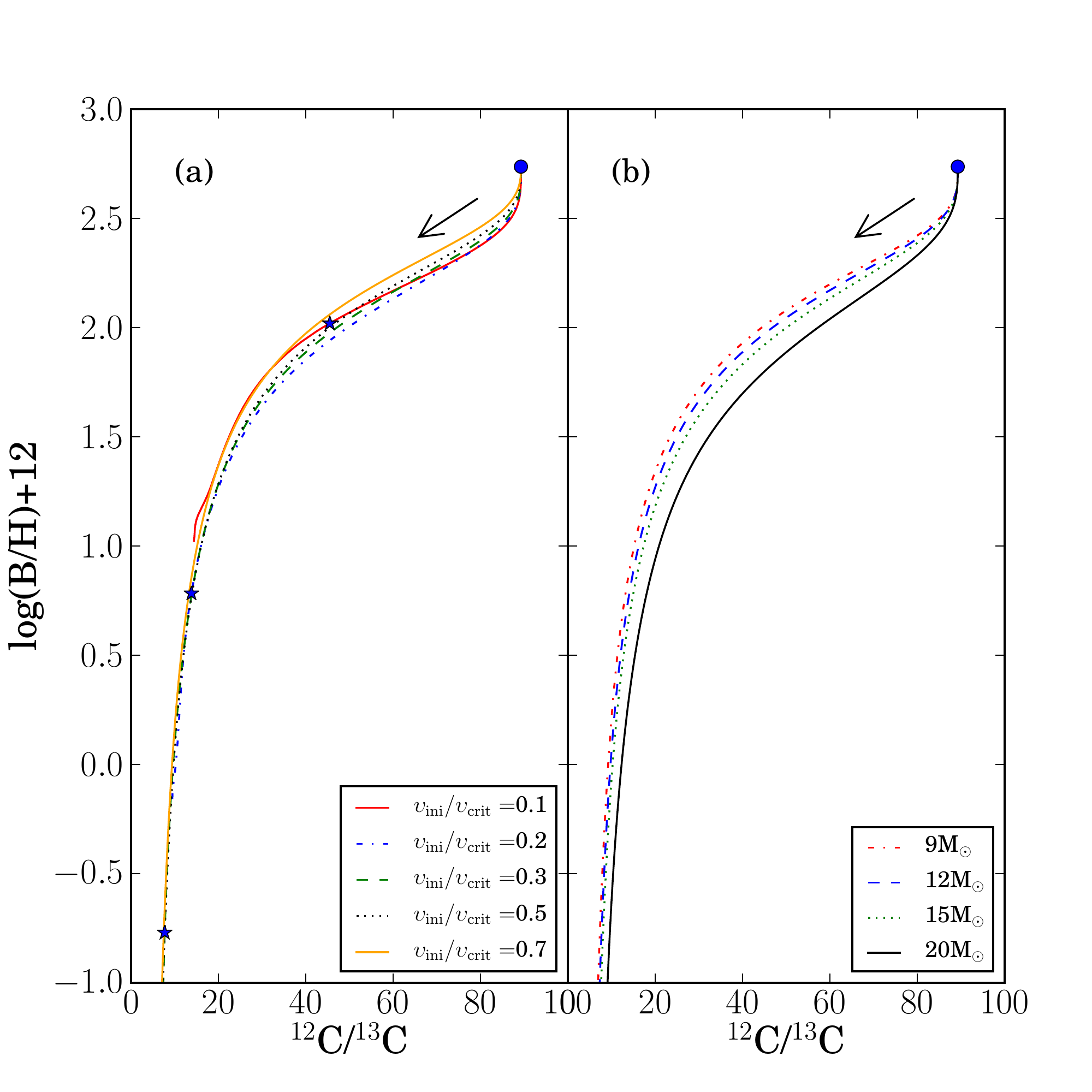}
 \caption{Boron versus $^{12}$C to $^{13}$C ratio for 12~M$_\odot$ models with different rotation velocities (a) and of 9, 12, 15, and 20~M$_\odot$ models with $\upsilon_{\rm ini}/\upsilon_{\rm{crit}}=$0.4 (b). In the left hand plot, the end of the MS of the slow rotators ($\upsilon_{\rm ini}/\upsilon_{\rm{crit}}=$0.1-0.3) is marked by a star symbol. The ZAMS position is indicated by a circle.}
 \label{fig:BvsC12C13}
\end{figure}

In Fig.~\ref{fig:BvsC12C13} the boron versus $^{12}$C$/^{13}$C ratio shows the same property as for boron versus nitrogen; i.e., the curve is almost independent of the parameters velocity and stellar mass in the investigated parameter range. The initial value for the $^{12}$C$/^{13}$C ratio is around 89 \citep{2003ApJ...591.1220L}. The ratio continuously decreases during the MS phase. On the surface, the $^{12}$C$/^{13}$C CNO-equilibrium value is only reached by the fastest rotators with a time-averaged equatorial velocity over $200$~km~s$^{-1}$ on the MS. 

\begin{figure}
 \includegraphics[width=0.485\textwidth]{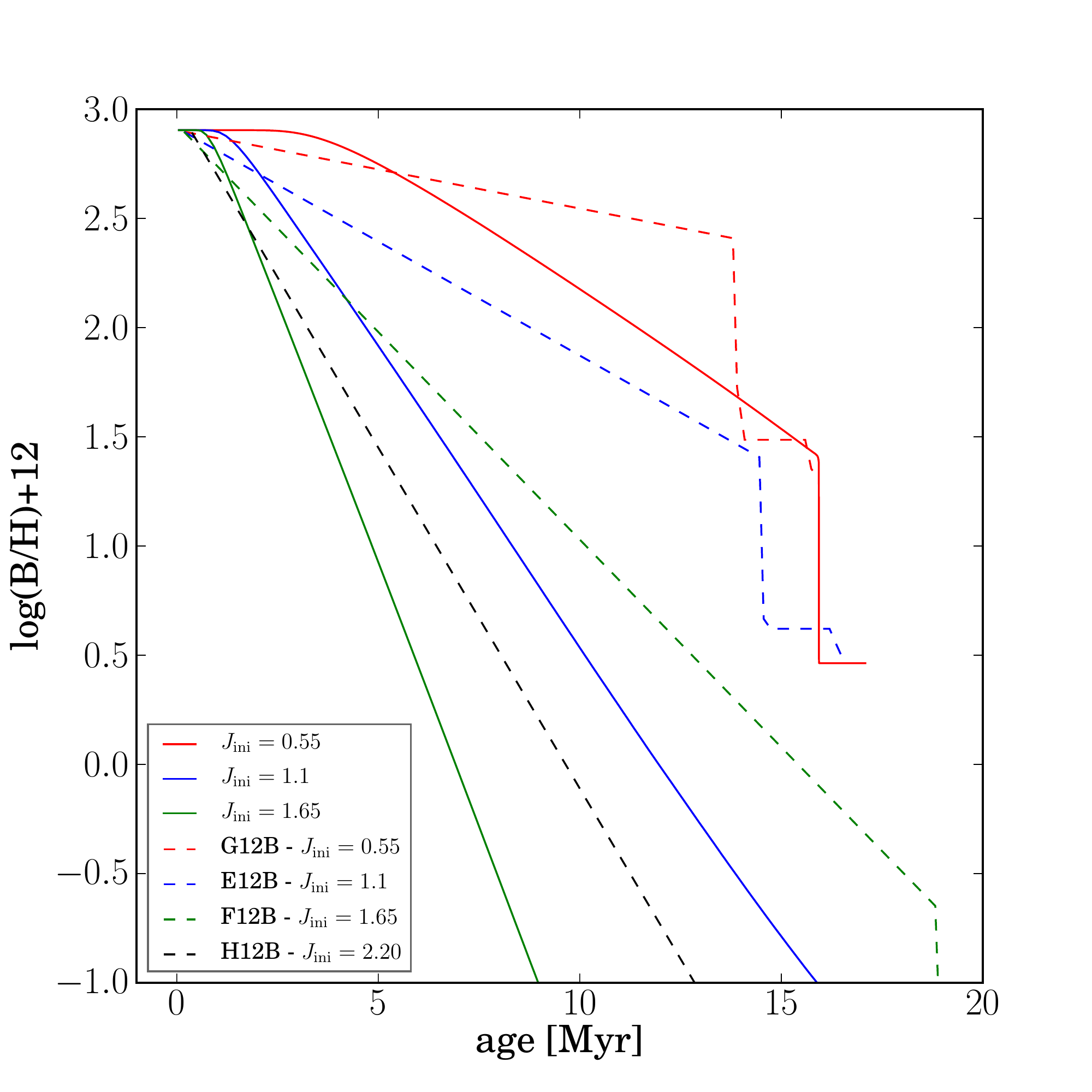}
 \caption{Boron versus time for 12~M$_\odot$ models with different rotation velocities, our models (solid lines) and 
 models from \citet{2000ApJ...544.1016H} (dashed lines) with comparable initial angular momentum $J_{\rm ini}~[10^{52}$~erg~s$]$. Both sets of models were calculated with Z=0.02.}
 \label{fig:logBvst_comparison}
\end{figure}

A comparison with the models of \citet{2000ApJ...544.1016H} is shown in Fig.~\ref{fig:logBvst_comparison}. For the purpose of comparison, we computed three 12~M$_\odot$ models with similar initial angular momentum and composition to the models of \citet{2000ApJ...544.1016H}. We see that the present models show more boron depletion at the end of the MS phase than the ones of \citet{2000ApJ...544.1016H}. The cause of this difference may be the different way of implementing the effects of rotation. An important difference is the way the advection of the angular momentum transport due to meridional currents is treated, as a diffusive process in the model by \citet{2000ApJ...544.1016H} and as an advective one in the present model. Also the counteractive effect of $\mu$-gradient on the shear diffusion is not treated in the same way in both models \citep[see][]{2000ApJ...544.1016H,2000A&A...361..101M}. We can also mention here that the models with solar like composition according to \citet{2005ASPC..336...25A}, i.e. with Z=0.014, show even stronger surface mixing. Lower metal content makes rotational mixing more efficient \citep{2001A&A...373..555M}, since the stars are more compact. Beyond these differences, we obtain here similar qualitative results to \citet{2000ApJ...544.1016H}, namely that boron depletion occurs much more rapidly than surface nitrogen enhancements.

\section{Comparison with the observations}
\label{sec:obs}
In Tables \ref{tab:obsdata}  and \ref{tab:abundobs}, the physical properties and the surface abundances of a selected sample of stars having boron determination are presented. Out of the available OB-type stars with boron detection from B III line at 2065.8 \AA, we chose only those with $T_{\rm eff}$ between 18'000 and 29'000~K. The B III 2065.8 \AA ~line strength has a plateau \citep{2002ApJ...565..571V} in this temperature range, making the B-determination more precise. All stars of this selection have narrow line spectra \citep[see e.g.][]{2001ApJ...548..429P}, which was a selection criterion to avoid blending effects. The $\upsilon\sin i$ is therefore low ($<$ 70~km~s$^{-1}$), meaning that either the equatorial velocity is low or that the star is seen nearly pole on.

\begin{figure}
  \includegraphics[width=0.485\textwidth]{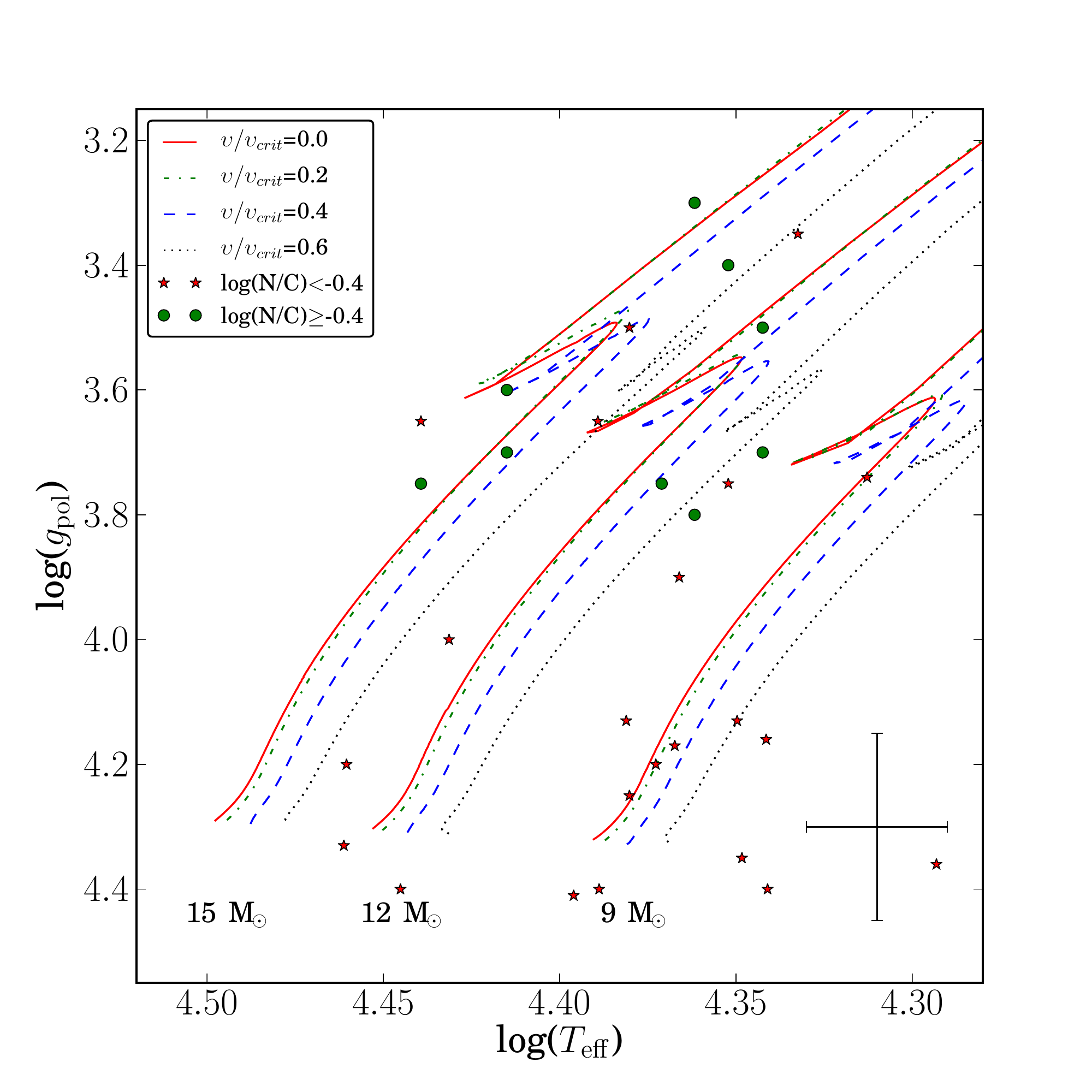}
  \caption{Evolutionary tracks of our models in the log$g_{\rm pol}$ versus log$T_{\rm eff}$ plane. For estimating the gravity of the theoretical track, we used the gravity at the pole. The gravities for the observed  stars are those deduced from spectroscopy. Different kinds of points are used for stars with various surface nitrogen over carbon ratios. The stars labelled with numbers correspond to the objects, discussed in the text. The typical error bars are depicted in the lower right hand corner.}
  \label{fig:gvsteff}
\end{figure}

The observations of boron in young massive stars show variations in log$\epsilon$(B) from 2.9 down to unobservable quantities below 1 \citep{2006ApJ...640.1039M,2002ApJ...565..571V,2001ApJ...548..429P}.  
Their positions in the log($g_{\rm pol}$) versus log($T_{\rm eff}$) plane is shown in Fig.~\ref{fig:gvsteff}. For plotting the tracks, we used the polar gravity because it is not affected much by rotation and is a good indicator of the evolutionary stage of the stars. The ``observed gravities'', deduced from spectroscopy, of course do not necessarily correspond to the polar ones. There is equality between these two quantities when the star is slowly rotating. For the fast rotators, the observed gravity can be lower than the polar one if the star is, for instance, seen equator on \citep[see the nice discussion of that topic in][]{2006ApJ...648..591H}. Thus some points in Fig.~\ref{fig:gvsteff} might be shifted towards higher values (downwards) if it were possible to deduce the polar gravity from the observations; however, this effect is only important for very fast rotators ($\upsilon_{\rm ini}/\upsilon_{\rm crit}>0.7$).

We also see that there are stars below the ZAMS, with gravities above 4.3. Obviously, the above inclination effect cannot be invoked here since this effect would still push these stars to greater log$g$. But with the current uncertainties, this difference is not significant (see the sizes of the error bars in Fig.~\ref{fig:gvsteff}). From Fig.~\ref{fig:gvsteff}, we see that the majority of the observed stars have initial masses between 9 and 15 M$_\odot$. We also see that most of the nitrogen enriched stars are found in the upper part of the MS band. This is consistent with the idea that these surface enrichments result from an evolutionary process.  

\begin{figure*}
 \includegraphics[width=\textwidth]{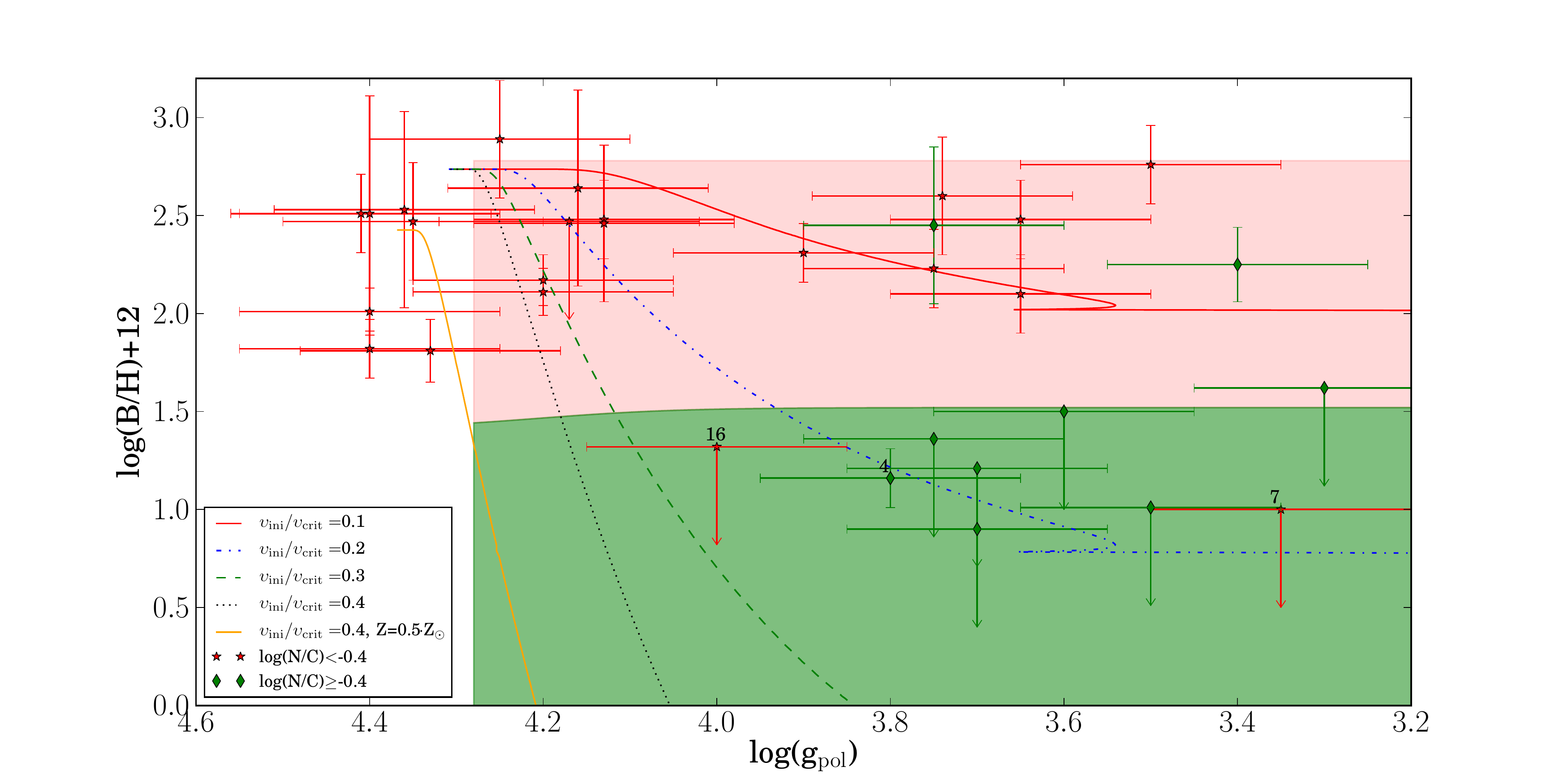}
 \caption{Surface boron abundances plotted versus polar gravity for 12~M$_\odot$ models with initial $\upsilon_{\rm ini}/\upsilon_{\rm crit}$ between 0 and 0.4. The orange continuous curve corresponds to the 12~M$_\odot$ with Z=0.007. Different kinds of points are used for stars with various surface N/C ratios. The labelled stars correspond to objects discussed in the text. The colour map depicts the same division in log(N/C) regimes as the observations going from low (top) to high (bottom) values.}
 \label{fig:logBvsgpol}
\end{figure*}

In Fig.~\ref{fig:logBvsgpol} the boron abundances with respect to hydrogen are plotted as a function of gravity. The colour map illustrates the regions where our 12~M$_\odot$ models with initial solar-like composition show different values of log(N/C). The upper area represents log(N/C)$<-0.4$ (red/light grey) and the lower one log(N/C)$\ge-0.4$ (green/dark grey). We see that most of the stars are accounted for well by the models with solar composition and slow-to-intermediate rotation rates ($\langle \upsilon_{\rm eq}\rangle=$0-100km~s$^{-1}$), assuming that the stars with highest gravities show their initial composition. The group of stars with log$\epsilon$(B) below about 1.7 are all more evolved stars, which is consistent with mixing processes occurring during the course of their evolution. In that diagram, stars with different N/C surface abundance ratios are plotted with different symbols. We see that, in agreement with models, most of the non-depleted boron stars show no nitrogen enrichment, and most of the boron depleted stars show nitrogen enrichments. This is indeed encouraging and can be taken as support for the mixing scenario.

\begin{figure*}
 \includegraphics[width=0.95\textwidth]{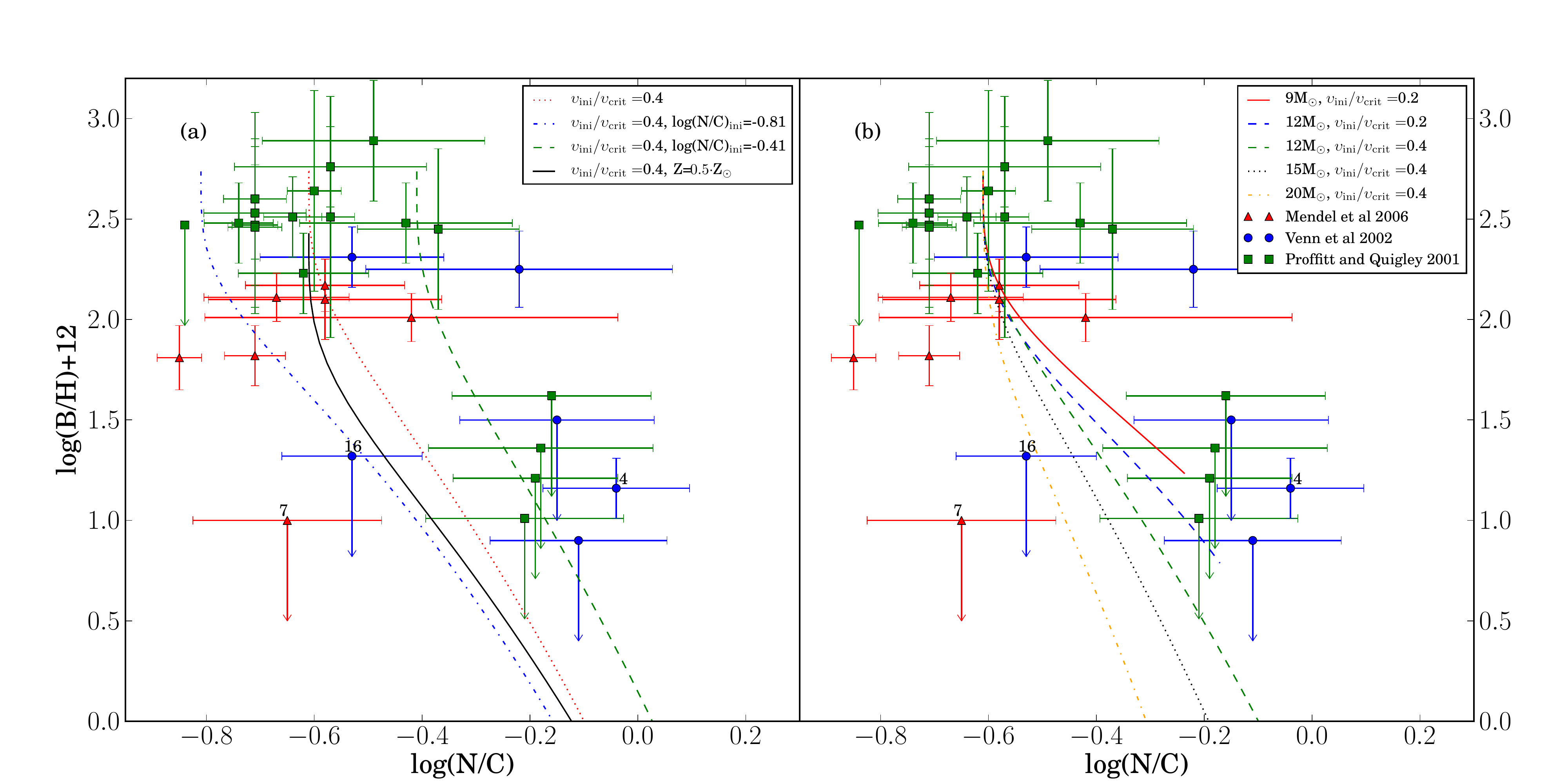}
 \caption{Boron versus nitrogen over carbon for the observations and in (a) 12M$_\odot$ models with different initial compositions and in (b) models with different masses and low-to-intermediate velocities to show the possible spread in MS evolution. The labelled stars correspond to objects that challenge the present theoretical predictions (see text).}
 \label{fig:logBvslogNC}
\end{figure*}

In Fig.~\ref{fig:logBvsgpol}, we see that two nitrogen-enriched stars (green diamonds) appear in the red area where models predict no or small nitrogen enrichment. But those stars have large error bars on their N/C, indicating that they may be normal N-rich stars (see also stars in Fig.~\ref{fig:logBvslogNC} with log$\epsilon$(B)$>$2 and log(N/C)$>-0.4$), therefore we shall not discuss them further in the present work.

One also sees that two stars are B-depleted while showing no nitrogen enrichment (stars 7 and 16). Such stars can be explained if the timescale for boron depletion is much shorter than the timescale for the surface nitrogen enrichment. Present models, even those rotating very fast, have difficulties reproducing the surface abundances of the two stars belonging to this category. Probably here, another process than the processes studied in the present paper must be invoked. Binary mass transfer usually occurs in case B\footnote{Mass transfer during the transition from MS to the red-giant phase.}, allowing only transfer of B-depleted and N-enriched material. But in the less common case A\footnote{Roche lobe overflow already during the main sequence of the donor.}, we might imagine transfer of B-depleted, but not yet N-enriched material. In close binaries, some mixing might also occur through tidal mixing, a process that has not yet been explored so far whether from an observational point of view or from theory. Another possibility would be that these stars present a strong differential rotation at the surface, triggered by some (magnetic?) braking mechanism of the surface layers. This would in turn trigger efficient mixing through shear instabilities in the layers with a strong differential rotation, explaining the rapid depletion of boron without any significant enhancement of nitrogen.

As just discussed, the bulk of the observations can be explained by our tracks with an average rotation on the MS between 0 and 100~km~s$^{-1}$. Looking at Fig.~\ref{fig:logBvsveq}, we see that the most B-depleted stars can be explained by 12 M$_\odot$ stellar models with a velocity on the MS superior to about 90~km~s$^{-1}$ and an inclination inferior or equal to 30~degrees\footnote{The inclination angle fix the position of the boron depletion curve on the MS at a given $\upsilon \sin i$, while the rotation velocity governs the amplitude of the depletion.}.

What would be interesting is to know the fraction of stars showing a $\upsilon\sin i$ inferior to $\upsilon_1$ having a velocity $\upsilon$ superior to a given limit, $\upsilon_2$. To compute such a fraction, one has to apply the equation
\begin{equation}
   P(\upsilon \sin i <   \upsilon_1\  {\rm with}\  \upsilon >  \upsilon_2)=
   \frac{\int\limits_{\upsilon_2}^{\upsilon_{\rm max}}\int\limits_0^{\varphi} f(\upsilon)\sin \alpha  d\upsilon d\alpha}
   {\int\limits_0^{\upsilon_{\rm max}}\int\limits_0^{\varphi} f(\upsilon)\sin \alpha  d\upsilon d\alpha},
\label{eq:dis}
\end{equation}  
where $f(\upsilon)$ is the  velocity distribution function, $\upsilon_{\rm max}$ the upper limit of the rotation velocities, and $\varphi$ the inclination angle between the rotation axis and the line of sight (equal to $\pi/2$ when the axis of rotation is perpendicular to the line of sight). $\varphi=\pi/2$ when $\upsilon\le\upsilon_1$ and $\varphi=arcsin(\upsilon_1/\upsilon)$ when $\upsilon \ge \upsilon_1$, so that $\upsilon\sin i\le\upsilon_1$. We suppose an isotropic distribution of inclination angles, so that the probability of having an inclination angle between $\alpha$ and $\alpha + d\alpha$ is proportional to $\sin \alpha d\alpha$. The denominator of Eq.~\ref{eq:dis} is proportional to the number of stars with $\upsilon\sin i\le\upsilon_1$, whereas the numerator counts the subset of these stars with $\upsilon >  \upsilon_2$.

For $\upsilon_1$ we adopted 50~km~s$^{-1}$, which is appropriate for the investigated sample, a value equal to 100~km~s$^{-1}$ was considered for $\upsilon_2$ (see above), since this is the velocity required to reproduce the strongest B-depletion. The value of $\upsilon_{\rm max}$ has been chosen as equal to 400 km~s$^{-1}$~\footnote{The results only marginally depend on the upper integration limit as long $\upsilon_{\rm max}\apprge$400km~s$^{-1}$.}. We assumed a Gaussian velocity distribution as proposed by \citet{2006A&A...457..265D} with the updated parameters, $\upsilon_0=$225~km~s$^{-1}$ and $\Delta\upsilon=$145~km~s$^{-1}$, from \citet{2009A&A...504..211H}. 

We find that 36\% of stars with $\upsilon \sin i <$ 50 km s$^{-1}$ have velocities $\upsilon$ superior to 100 km s$^{-1}$. Considering that our model 12~M$_\odot$ with $\upsilon_{\rm ini}/\upsilon_{\rm crit}=0.2$ only reaches the upper limits for B-depletion at the very end of its MS, $\upsilon_2=$100~km~s$^{-1}$ might seem too optimistic. But with stricter limits, such as $\upsilon_2=$125 or even 150~km~s$^{-1}$, the probability is still 29\% and 23\%, respectively. This fits the 29\% (9 of 31 stars) B-depleted stars in the sample. 

If we take only the sample with $\upsilon\sin i<$20~km~s$^{-1}$ ($\upsilon_1=$20~km~s$^{-1}$), then we get $P=0.21$, i.e. 21\% whereas in the observational sample 25\% (3 out of 12 stars) are B-depleted (compare Fig.~\ref{fig:logBvsveq}b), which  is also in good agreement for the size of this sub-sample. We can therefore conclude that the inclination effect is likely to play a role in the observational sample here and that the statistical properties fit our models of single rotating stars. 

In Fig.~\ref{fig:logBvslogNC}, we have plotted the observations of the B/H versus N/C ratio, as well as the evolutionary tracks obtained in the present work. This diagram tests whether the concomitant changes of boron and nitrogen are reproduced by the rotating stellar models. In Fig.~\ref{fig:logBvslogNC}a we plotted models with $\upsilon_{\rm ini}/\upsilon_{\rm crit}=0.4$ to show the typical B-N/C relation. 
We see that these models could explain a large fraction of the observations, but they correspond to an average velocity on the MS of around 180~km~s$^{-1}$, while  the observational sample mainly consists of slow rotators ($\upsilon\sin i\le$50~km~s$^{-1}$). On the other hand, in panel b of Fig.~\ref{fig:logBvslogNC},  models with lower velocities are shown. We see that they would provide a good fit to the points with boron abundances between 0.9 and 1.5 and N/C ratios of the order of $-0.2$. These models have difficulty explaining stars that are more depleted in boron and more enriched in nitrogen. 

The two stars 7 and 16  (see Fig.~\ref{fig:logBvslogNC}) present no or little surface nitrogen enrichment and are strongly boron-depleted. The range of values spanned by the different initial mass and initial velocity models barely explain these values, therefore these stars challenge our models even though the difference in N/C is within 2-$\sigma$ of the models at the upper limit. These two stars were also found to be a problem for single-star models by other authors \citep{2008A&A...481..453M,2006ApJ...640.1039M,2002ApJ...565..571V}.

Apart from these few difficult cases, the above comparisons are very encouraging on the whole and support the predictions of the rotating models. At this stage it is also interesting to note that star 7, which cannot be reproduced by the present models, is a spectroscopic binary with a period of 9.5 days \citep{2004A&A...424..727P}, so it might well be that this star was slowed down by spin-orbit coupling in the course of its MS evolution. Such a braking mechanism would also slow down or even stop the N-enrichment process.

Are there any other indications that the changes in surface abundances we observe here are linked to rotation? To answer that question, measurements of surface velocities would be welcome. At the moment only a few stars (5) have estimates of their surface velocity (see Table~\ref{tab:obsdata}). These stars are plotted in Fig.~\ref{fig:logBvsveq}. Let us make a few comments on each of them.

\begin{itemize}
\item Stars 3 and 4: for those stars, the present rotating models show too high surface velocity for their boron surface abundances. In other words, in Fig.~\ref{fig:logBvsveq}a, the theoretical MS tracks that go through the observed B/H ratio {\it during the MS phase} are too much to the right. This may indicate that mixing in those stars is more efficient than predicted by the present models\footnote{In case those stars were post MS stars, which is not confirmed by their present observed gravities (see Fig.~\ref{fig:logBvsgpol}), then there would be no difficulty for the present models to explain those stars.}. This is supported by the high N surface content and that they are still on the MS. Models computed with a lower initial metal content may help to improve the situation (see yellow track and light red shaded area in Fig.~\ref{fig:logBvsveq}a). Indeed, as can be seen looking at the track for the 12 M$_\odot$, $\upsilon/\upsilon_{\rm crit}=0.4$, Z=0.5 Z$_\odot$, the MS band extends well into the post-MS regions defined by the more metal-rich models. In that case there would be no difficulty at least reproducing star 3. But the iron content of these stars is not significantly different from the solar value. In general, most stars in the sample have Fe surface abundance consistent with the solar value\footnote{Stars 5 and 8 have a log$\epsilon$(Fe) differing by more than 2 $\sigma$ but do not show B-depletion.}. Star 4 is the most extreme case in showing B-depletion while rotating slowly and is therefore positioned in the shaded area of Fig.~\ref{fig:logBvsveq}a. It cannot be explained even with a low-metallicity track.

\item Stars 6, 24 and 30:  for these stars, the present theoretical models present characteristics that agree with the observations. Looking at Fig.~\ref{fig:logBvsveq}a, star 6 might be interpreted as a slowly rotating star at the beginning of its evolution, but its gravity indicates it is an evolved star that is rather at the end of the MS phase, if not beyond. Thus it seems that here the mixing was less efficient than shown by the present models. But the $\sigma$-error of boron is comparable to the boron depletion expected at this low rotation rate, so at the moment, this observation is not very constraining and is compatible with the present models.
\end{itemize}

Except for star 4, present models explain surface abundances, velocities and gravities simultaneously. The number of stars is, however, small, and it would be interesting to obtain more velocity measurements to set such a conclusion on firmer ground. 

If we look at the projected velocities in Fig.~\ref{fig:logBvsveq}b, it appears that all B-depleted stars are in the shaded regions, {\it i.e.} beyond the main sequence phase, which does not appear to be consistent with the log $g$ values (see Fig.~\ref{fig:logBvsgpol}). This difficulty might be resolved by invoking the following reasons.
\begin{enumerate}
\item Initially rapidly rotating stars were suddenly slowed down at a given point by, for example, binary interaction. 
\item The efficiency of mixing is underestimated, so that stars can be both boron-depleted and rotating at low velocities. Rotation is measured only at the surface, while the shear turbulence, which governs the changes in the surface abundances in rotating models depends on the gradient of the velocity inside the star. Effects like magnetic breaking could trigger mixing in the outer layers and slow the surface down.
\item The stars have low inclination angles (see the effect illustrated in Fig \ref{fig:logBvsveq}). 
\end{enumerate}
The last case was discussed at the start of this section and, from a statistical point of view, may indeed explain part of the discrepancy. Point 2 seems to be required at least for some stars (like 7 and 16) only on the basis of their position in the boron versus nitrogen plane. The impact of binarity still remains difficult to assess in part because this effect produces similar changes in the surface abundances to those expected from rotation, at least for those stars showing boron depletion and nitrogen enrichment\footnote{Models with intermediate-to-fast rotation ($\upsilon/\upsilon_{\rm crit}\ge0.3$) show comparable boron depletion ($>3$~dex) to binary mass transfer.}. 

A way to disentangle rotation and binary effect would be to look at the presence or not of a correlation between surface velocity and surface enrichments. In case binarity effects dominate, no correlation is expected between stellar rotation rate and boron depletion \citep{2010IAUS..268..411L}, while rotational mixing would lead to stronger depletion for faster rotation rates (for the same mass, metallicity, and age). In Fig.~\ref{fig:logBvsveq} no correlation can be seen. This might look like favouring binary mixing. But for such a conclusion to be valid, one needs to correct for the inclination effects and one should be sure that stars of similar initial masses and ages form the bulk of the sample\footnote{Mixing efficiency depends not only on the rotation velocity alone but also on the age, mass, and metallicity.}, which is not the case here, looking at Fig.~\ref{fig:gvsteff}. From the present data, therefore, we have difficulty concluding about the nature of the process at work. Rotation does appear to provide a reasonable process for explaining at least part of the stars, but binarity can also play a role. 

From an observational point of view, we may conclude that, to take a step further, the following informations would be needed:
\begin{itemize}
\item The boron abundances on the surface of stars with a high $\upsilon\sin i$. This is probably a real challenge to observe, but since $\upsilon\sin i$ is a lower limit to $\upsilon$, any high values imply a fast rotator. Unless the observed stars are at the very beginning of the main sequence, rotational models would predict that all these stars should be boron-depleted.
\item The abundances of $^3$He and the values of the $^{12}$C/$^{13}$C ratio in order to test the predictions of the present models shown in Figs. \ref{fig:logBvslogHe3} and \ref{fig:BvsC12C13}. Of course it would be interesting to have predictions of models invoking binary mass transfer and/or tidal mixing for comparison.
\item Observations in star clusters or of eclipsing binaries, since they would allow stars of the same age to be compared, with those for which some (more) precise indications on the mass can be obtained. 
\item Asteroseismological data could probe the way rotation varies with depth, but it is only feasable for very slow rotators. Asteroseismology is  also able, together with models, to provide some information on the previous rotational history of the star \citep{2008IAUS..250..237A}. This might offer very interesting hints to the processes at work here.
\end{itemize}

\section{Conclusions}
\label{sec:conclusion}
We implemented the Basel reaction network into the Geneva stellar evolution code (GENEC), which treats the meridional circulation as an advection process. This allowed us to predict the surface evolution of light elements in massive stars. 
We obtained the following results.
\begin{itemize}
\item The boron depletion is stronger than in models of \citet{2000ApJ...544.1016H}. This is probably because we account for the effect of meridional currents on the transport of angular momentum as an advective process and not as a diffusive one, along with accounting for the effects of mean molecular weight gradients on the efficiency of rotational diffusion, which is different in both models.

\item We present expected correlations based on rotating stellar models between boron surface abundances and the surface abundances of $^3$He, and the surface number ratios  $^{12}$C/$^{13}$C. Further observations will be able to check whether these correlations are present in real stars.

\item We confirm the general conclusion obtained by \citet{2006ApJ...640.1039M} and \citet{2002ApJ...565..571V} that rotational mixing can account for most of the observations of the boron and nitrogen surface abundances.

\item We confirm the existence of challenging cases that do not fit well in the present scenario: stars 4 and 7, which do appear to present a much more rapid B-depletion than presently allowed by the models.

\item Even though our models can reproduce most of the observations, the current uncertainties do not allow us to draw a firm conclusion about the questions whether our models mix enough and whether another surface mixing mechanism has to be invoked. We proposed some possibilities for further observations that would help clarifying the situation.

\end{itemize}

\begin{acknowledgements}
Many thanks go to Olivier Schnurr for his helpful comments. This project was supported by the Swiss National Science Foundation. 
\end{acknowledgements}

\begin{table*}
\caption{Stellar parameters}            
 \begin{tabular}{rlccccccccr}
\hline\hline \\[-3pt]
No & Star 	& Cluster     & Binary  & log$g^a$ & $T_{eff}^b$ & $\upsilon\sin i$ & Ref. & $\upsilon_{\rm eq}$ & Ref. & Mass$^c$ \\
   &		&	      &         &      &       & [km~s$^{-1}$] &   & [km~s$^{-1}$] && [M$_\odot$] \\
\hline\\[-5pt]
1 & BD +56$^\circ$576&$\chi$ Per& Eclipse& 3.40& 22500 & 17       &  1  &\dots    &   & 13.5 \\  
2 & HD 886   	& \dots       & \dots   & 3.75 & 22500 & 10$\pm$1 &  2  &\dots    &   & 10.7 \\  
3 & HD 3360   	& CasTau OB1  & \dots   & 3.70 & 22000 & 19$\pm$1 &  3  & 55$\pm$28 & 7 & 10.6 \\ 
4 & HD 16582   	& CasTau OB1  & \dots   & 3.80 & 23000 & 14$\pm$1 &  2  &14/28 & 8 & 10.7 \\ 
5 & HD 22951   	& Per OB      & Visual  & 4.40 & 27900 & 23       &  4  &\dots&   & 11   \\ 
6 & HD 29248   	& Ori OB1     & \dots   & 3.75 & 23500 & 36$\pm$3 &  2  &  6  & 9 & 11.5 \\ 
7 & HD 30836   	& Ori OB1     & Spect.  & 3.35 & 21500 & 43$\pm$3 &  5  &\dots&   & 13.0 \\ 
8 & HD 34816   	& Ori runaway & \dots   & 4.20 & 28900 & 35       &  4  &\dots&   & 13.5 \\ 
9 & HD 35039   	& Ori OB1a    & Spect.  & 3.74 & 20600 &  4       &  6  &\dots&   &  9.0 \\ 
10& HD 35299   	& Ori OB1a    & \dots   & 4.25 & 24000 &  8       &  6  &\dots&   &  9.1 \\ 
11& HD 35337   	& Ori OB1c    & \dots   & 4.20 & 23600 & 15       &  4  &\dots&   &  9.0 \\ 
12& HD 35468   	& Ori OB1     & \dots   & 3.50 & 22000 & 51$\pm$4 &  5  &\dots&   & 11.8 \\ 
13& HD 36285   	& Ori OB1c    & \dots   & 4.40 & 21900 & 10       &  4  &\dots&   & $<$9 \\ 
14& HD 36351   	& Ori OB1a    & Visual  & 4.16 & 22000 & 42       &  6  &\dots&   & $<$9 \\ 
15& HD 36430   	& Ori OB1c    & \dots   & 4.36 & 19600 & 26       &  6  &\dots&   & $<$9 \\ 
16& HD 36591   	& Ori OB1b    & Visual  & 4.00 & 27000 & 16$\pm$2 &  5  &\dots&   & 12.8 \\ 
17& HD 36629   	& Ori OB1c    & \dots   & 4.35 & 22300 &  4       &  6  &\dots&   & $<$9 \\ 
18& HD 36959   	& Ori OB1c    & \dots   & 4.41 & 24900 & 11       &  6  &\dots&   &  9   \\ 
19& HD 36960   	& Ori OB1c    & Visual  & 4.33 & 28900 & 33       &  4  &\dots&   & 12   \\ 
20& HD 37209   	& Ori OB1c    & \dots   & 4.13 & 24000 & 50       &  6  &\dots&   &  9.7 \\ 
21& HD 37356   	& Ori OB1c    & \dots   & 4.13 & 22400 & 23       &  6  &\dots&   & $<$9 \\ 
22& HD 37481   	& Ori OB1c    & \dots   & 4.17 & 23300 & 67       &  6  &\dots&   & $<$9 \\ 
23& HD 37744   	& Ori OB1b    & \dots   & 4.40 & 24500 & 39       &  6  &\dots&   &  9   \\ 
24& HD 44743   	& \dots       & \dots   & 3.50 & 24000 & 23$\pm$2 &  2  & 31$\pm$5  & 10& 14.5 \\ 
25& HD 46328   	&  Coll. 121  & \dots   & 3.75 & 27500 & 10$\pm$2 &  2  &\dots&   & 15.5 \\ 
26& HD 50707   	&  Coll. 121  & \dots   & 3.60 & 26000 & 45$\pm$3 &  5  &\dots&   & 15.6 \\ 
27& HD 52089   	& \dots       & \dots   & 3.30 & 23000 & 28$\pm$2 &  5  &\dots&   & 15.3 \\
28& HD 111123	& Sco Cen     & \dots   & 3.65 & 27500 & 48$\pm$3 &  5  &\dots&   & 16.6 \\
29& HD 205021   & Cep OB1     & Spect.  & 3.70 & 26000 & 29$\pm$2 &  2  &\dots&   & 14.5 \\ 
30& HD 214993   & Lac OB1     & \dots   & 3.65 & 24500 & 42$\pm$4 &  2  & 45  & 11& 13.5 \\ 
31& HD 216916   & Lac OB1     & Eclipse & 3.90 & 23200 & 13       &  1  &\dots&   &  9.7 \\ 
\hline\\[-5pt]
 \end{tabular}
\\
$^a$The typical 1-$\sigma$ error for $\log g$ is between 0.15 and 0.2 \citep[see e.g][]{2008A&A...481..453M}. \\
$^b$The typical 1-$\sigma$ error for $T_{eff}$ is about 1000~K \citep[see e.g][]{2008A&A...481..453M}.\\
$^c$The masses of the stars are only rough estimates from the comparison of the models and the stellar positions in the log$g$-log$T_{\rm eff}$ diagram under the assumption of slow rotation. \\
References: [1] \citet{2002ApJ...565..571V}; [2] \citet{2006A&A...457..651M};  [3] \citet{2007CoAst.150..183B}; [4] \citet{2006ApJ...640.1039M}; [5] \citet{2008A&A...481..453M}; [6] \citet{2001ApJ...548..429P}; [7] \citet{2003A&A...406.1019N}; [8] \citet{2006ApJ...642..470A}; [9] \citet{2004MNRAS.350.1022P}; [10] \citet{2006A&A...459..589M}; [11] \citet{1996A&A...314..115A}\\
\label{tab:obsdata}
\end{table*}

\begin{table*}
\caption{Surface abundances}            
 \begin{tabular}{rlcccccccc}
\hline\hline \\[-3pt]
No & Star 	& log$\epsilon$(C) & log$\epsilon$(N) & log(N/C) & Ref. & log$\epsilon$(Fe) & Ref.  & log$\epsilon$(B) & Ref.  \\
\hline\\[-5pt]
1 & BD +56$^\circ$576	& 7.84$\pm$0.18 & 7.62$\pm$0.22 &  -0.22$\pm$0.28 & 1$^a$ & 7.34$\pm$0.17 &  1  & 2.25$\pm$0.19 &  1  \\  
2 & HD 886   		& 8.20$\pm$0.05 & 7.58$\pm$0.11 &  -0.62$\pm$0.12 & 2     & 7.25$\pm$0.16 &  2  & 2.23$\pm$0.20 &  6  \\  
3 & HD 3360   		& 8.16$\pm$0.08 & 7.97$\pm$0.13 &  -0.19$\pm$0.15 & 3     & 7.31$\pm$0.16 &  3  & $<$1.21       &  6  \\ 
4 & HD 16582   		& 8.09$\pm$0.08 & 8.05$\pm$0.11 &  -0.04$\pm$0.14 & 2     & 7.32$\pm$0.18 &  2  & 1.16$\pm$0.15 &  1  \\ 
5 & HD 22951   		& 8.11$\pm$0.21 & 7.69$\pm$0.32 &  -0.42$\pm$0.38 & 1$^a$ & 7.03$\pm$0.10 &  4  & 2.01$\pm$0.12 &  4  \\ 
6 & HD 29248   		& 8.24$\pm$0.12 & 7.87$\pm$0.09 &  -0.37$\pm$0.15 & 2     & 7.36$\pm$0.19 &  2  & 2.45$\pm$0.40 &  6  \\ 
7 & HD 30836   		& 8.19$\pm$0.09 & 7.54$\pm$0.15 &  -0.65$\pm$0.18 & 5     & 7.11$\pm$0.18 &  5  & $<$1.00       &  4  \\ 
8 & HD 34816   		& 8.17$\pm$0.07 & 7.59$\pm$0.13 &  -0.58$\pm$0.15 & 1$^a$ & 7.12$\pm$0.11 &  4  & 2.17$\pm$0.13 &  4  \\ 
9 & HD 35039   		& 8.36$\pm$0.03 & 7.65$\pm$0.05 &  -0.71$\pm$0.06 & 1$^a$ & 7.24	  &  6  & 2.60$\pm$0.30 &  6  \\ 
10& HD 35299   		& 8.19$\pm$0.19 & 7.70$\pm$0.08 &  -0.49$\pm$0.21 & 1$^a$ & 7.19	  &  6  & 2.89$\pm$0.30 &  6  \\ 
11& HD 35337   		& 8.31$\pm$0.09 & 7.64$\pm$0.10 &  -0.67$\pm$0.13 & 1$^a$ & 7.38$\pm$0.10 &  4  & 2.11$\pm$0.12 &  4  \\ 
12& HD 35468   		& 8.11$\pm$0.09 & 7.90$\pm$0.16 &  -0.20$\pm$0.19 & 5     & 7.23$\pm$0.14 &  5  & $<$1.01       &  6  \\ 
13& HD 36285   		& 8.48$\pm$0.04 & 7.77$\pm$0.04 &  -0.71$\pm$0.06 & 1$^a$ & 7.23$\pm$0.09 &  4  & 1.82$\pm$0.15 &  4  \\ 
14& HD 36351   		& 8.28$\pm$0.04 & 7.68$\pm$0.03 &  -0.60$\pm$0.05 & 1$^a$ & 7.28          &  6  & 2.64$\pm$0.50 &  6  \\ 
15& HD 36430   		& 8.38$\pm$0.03 & 7.67$\pm$0.09 &  -0.71$\pm$0.09 & 1$^a$ & 7.54	  &  6  & 2.53$\pm$0.50 &  6  \\ 
16& HD 36591   		& 8.19$\pm$0.05 & 7.66$\pm$0.12 &  -0.53$\pm$0.13 & 5     & 7.32$\pm$0.19 &  5  & $\le$1.32     &  1  \\ 
17& HD 36629   		& 8.32$\pm$0.03 & 7.61$\pm$0.03 &  -0.71$\pm$0.04 & 1$^a$ & 7.39	  &  6  & 2.47$\pm$0.30 &  6  \\ 
18& HD 36959   		& 8.37$\pm$0.02 & 7.73$\pm$0.05 &  -0.64$\pm$0.05 & 1$^a$ & 7.29	  &  6  & 2.51$\pm$0.20 &  6  \\ 
19& HD 36960   		& 8.39$\pm$0.01 & 7.54$\pm$0.04 &  -0.85$\pm$0.04 & 1$^a$ & 7.22$\pm$0.10 &  4  & 1.81$\pm$0.16 &  4  \\ 
20& HD 37209   		& 8.29$\pm$0.04 & 7.55$\pm$0.05 &  -0.74$\pm$0.06 & 1$^a$ & 7.32	  &  6  & 2.48$\pm$0.20 &  6  \\ 
21& HD 37356   		& 8.41$\pm$0.03 & 7.70$\pm$0.04 &  -0.71$\pm$0.05 & 1$^a$ & 7.32	  &  6  & 2.46$\pm$0.40 &  6  \\ 
22& HD 37481   		& 8.39          & 7.55$\pm$0.02 &  -0.84          & 1$^a$ & 7.40	  &  6  & $<$2.47       &  6  \\ 
23& HD 37744   		& 8.37$\pm$0.02 & 7.80$\pm$0.04 &  -0.57$\pm$0.04 & 1$^a$ & 7.36	  &  6  & 2.51$\pm$0.60 &  6  \\ 
24& HD 44743   		& 8.16$\pm$0.11 & 7.59$\pm$0.14 &  -0.57$\pm$0.18 & 2     & 7.17$\pm$0.19 &  2  & 2.76$\pm$0.20 &  6  \\ 
25& HD 46328   		& 8.18$\pm$0.12 & 8.00$\pm$0.17 &  -0.18$\pm$0.21 & 2     & 7.30$\pm$0.22 &  2  & $<$1.36       &  6  \\ 
26& HD 50707   		& 8.18$\pm$0.10 & 8.03$\pm$0.15 &  -0.15$\pm$0.19 & 5     & 7.23$\pm$0.19 &  5  & $\le$1.50     &  1  \\ 
27& HD 52089   		& 8.09$\pm$0.12 & 7.93$\pm$0.14 &  -0.16$\pm$0.19 & 5     & 7.16$\pm$0.15 &  5  & $<$1.62       &  6  \\
28& HD 111123		& 8.04$\pm$0.10 & 7.61$\pm$0.17 &  -0.43$\pm$0.20 & 5     & 7.23$\pm$0.24 &  5  & 2.48$\pm$0.20 &  6  \\
29& HD 205021  	 	& 8.02$\pm$0.10 & 7.91$\pm$0.13 &  -0.11$\pm$0.17 & 2     & 7.24$\pm$0.23 &  2  & $\le$0.90     &  1  \\ 
30& HD 214993   	& 8.22$\pm$0.12 & 7.64$\pm$0.18 &  -0.58$\pm$0.22 & 2     & 7.30$\pm$0.20 &  2  & 2.10$\pm$0.20 &  4  \\ 
31& HD 216916   	& 8.17$\pm$0.13 & 7.64$\pm$0.11 &  -0.53$\pm$0.17 & 1$^a$ & 7.66$\pm$0.14 &  1  & 2.31$\pm$0.15 &  1  \\ 
\hline\\[-5pt]
 \end{tabular}
\\
$^a$N and C values are from the compilation in \citet{2002ApJ...565..571V} where most N and C values are corrected values originally from \citet{1992ApJ...387..673G} and \citet{1994ApJ...426..170C}.\\
References: the numbers correspond to the same references as in Table \ref{tab:obsdata}.\\
\label{tab:abundobs}
\end{table*}

\end{document}